\documentclass[twoside,11pt]{article}
\usepackage{blindtext}

\usepackage{jmlr2e}
\usepackage{amsmath,amssymb}
\usepackage{hyperref}
\usepackage{newtxtext}
\usepackage{multirow}
\usepackage[subscriptcorrection]{newtxmath}
\usepackage{color}
\graphicspath{{./art/}}
\usepackage[ruled,noend]{algorithm2e}
\SetKwRepeat{Do}{do}{while}
\makeatletter
\renewcommand{\algocf@captiontext}[2]{#1\algocf@typo. \AlCapFnt{}#2} 
\def\@algocf@capt@plain{top}
\renewcommand{\algocf@makecaption}[2]{%
  \addtolength{\hsize}{\algomargin}%
  \sbox\@tempboxa{\algocf@captiontext{#1}{#2}}%
  \ifdim\wd\@tempboxa >\hsize
    \hskip .5\algomargin%
    \parbox[t]{\hsize}{\algocf@captiontext{#1}{#2}}
  \else%
    \global\@minipagefalse%
    \hbox to\hsize{\box\@tempboxa}
  \fi%
  \addtolength{\hsize}{-\algomargin}%
}

\newtheorem{assumption}{Assumption}

\newcommand{\var}{\textsf{Var}}

\renewcommand{\P}{\text{P}}
\newcommand{\iidsim}{\stackrel{\text{iid}}{\sim}}
\renewcommand{\d}{{\text{diff}}}

\newcommand{\covp}{\textsf{Cov}_\text{P}}
\newcommand{\epp}{\textsf{E}_\text{P}}
\newcommand{\varp}{\textsf{Var}_\text{P}}

\usepackage{lastpage}
\firstpageno{1}

\begin{document}

\title{Mitigating dimensionality effects with robust graph constructions for testing}

\author{\name Yejiong Zhu \email yjzhu@ucdavis.edu \\
       \addr Department of Statistics\\
       University of California, Davis\\
       Davis, CA, USA
       \AND
       \name Hao Chen \email hxchen@ucdavis.edu \\
       \addr Department of Statistics\\
       University of California, Davis\\
       Davis, CA, USA}

\editor{}

\maketitle

\begin{abstract}
Dimensionality effects pose major challenges in high-dimensional and non-Euclidean data analysis. Graph-based two-sample tests and change-point detection are particularly attractive in this context, as they make minimal distributional assumptions and perform well across a wide range of scenarios. These methods rely on similarity graphs constructed from data, with $K$-nearest neighbor graphs and $K$-minimum spanning trees among the most effective and widely used. However, in high-dimensional and non-Euclidean regimes such graphs often produce hubs -- nodes with disproportionately high degrees -- to which graph-based methods are especially sensitive. To mitigate these dimensionality effects, we propose a robust graph construction that is far less prone to hub formation. Incorporating this construction substantially improves the power of graph-based methods across diverse settings. We further establish a theoretical foundation by proving its consistency under fixed alternatives in both low- and high-dimensional regimes. The effectiveness of the approach is demonstrated through real-world applications, including comparisons of correlation matrices for brain regions, gene expression profiles of T cells, and temporal changes in New York City taxi travel patterns.

\end{abstract}


\section{Introduction}\label{sec: 1}
Two-sample comparison is a fundamental problem in statistical analysis, where researchers evaluate whether the characteristics of subjects from two different groups follow the same distribution. While extensively studied in the context of univariate or low-dimensional data, the problem becomes more challenging with the growing complexity of modern datasets. In fields such as genomics, neuroscience, finance, and social sciences, data are increasingly high-dimensional or non-Euclidean, posing new difficulties for traditional approaches. Many contemporary applications involve such complex data structures, including the following:
\begin{enumerate}

    \item Genomics and Bioinformatics: Two-sample testing plays a vital role in identifying differentially expressed genes between groups, such as healthy individuals and those with a specific disease. These analyses often involve high-dimensional gene expression datasets where the number of genes vastly exceeds the number of samples {\citep{dudoit2002statistical,lister2009human,ritchie2015limma}}. Two-sample tests are also widely used in microbiome research to compare microbial composition, where each sample may contain hundreds to thousands of taxa, and shotgun metagenomic sequencing can result in tens of thousands of dimensions \citep{lozupone2005unifrac,lozupone2012diversity}. 

    \item Neuroscience: Two-sample testing is commonly used to detect differences in functional brain connectivity between groups {\citep{zalesky2010network,di2014autism}}, thereby identify changes in neural network patterns associated with  cognitive states or neurological disorders. For example, \cite{xu2023data} analyzed MRI brain images across conditions such as Parkinson’s and Alzheimer’s disease. Depending on the parcellation strategy, the dimensionality ranges from over $1,000$ to nearly $10,000$, while the number of subjects in both patient and control groups is often in the hundreds or fewer. 
    
    \item Social sciences: Two-sample testing is employed to examine differences in interaction networks, such as comparing social structures before and after interventions (e.g., policy changes, community programs, or digital platform rollouts). These comparisons often involve network-valued data, where inference relies on comparing structural properties of graphs \citep{krackhardt1993informal, jackson2002evolution,ghoshdastidar2017two}, or on testing differences in global metrics such as centrality, clustering, or homophily \citep{ borgatti2011analyzing, fujiwara2022network}.
    
\end{enumerate}

Formally, the problem can be written as testing $H_0:F_X = F_Y\text{ versus } H_a: F_X\neq F_Y$, based on independent samples $X_1,\cdots, X_m \iidsim F_X$ and $Y_{1},\cdots, Y_n \iidsim F_Y$. In high-dimensional and non-Euclidean regimes, parametric methods are often limited by their reliance on specific distribution assumptions, which may not hold in practice. Defining an appropriate parametric model for high-dimensional data is inherently challenging, making nonparametric or distribution-free methods more appealing unless prior knowledge about the underlying distribution is available.

In the nonparametric domain, numerous advancements have been made over the years. Among these, graph-based tests have gained prominence due to their strong performance and reliable control of type I error \citep{chen2017new,chen2018weighted,zhang2022graph,zhou2023new,zhu2024limiting}. These methods construct similarity graphs from the pooled sample and define test statistics from the connectivity patterns of the graph. Because they require only pairwise distances, they are naturally suited to high-dimensional and non-Euclidean data.   

A foundational contribution in this line of work is due to \cite{friedman1979multivariate}, which proposed a procedure based on the minimum spanning tree (MST) of the pooled sample -- that is, the tree connecting all observations with minimal total edge lengths -- and defined the test statistic as the number of edges connecting observations from different groups. For clarity, we will refer to this method as the \emph{original edge-count test} (OET). This was also the first practical nonparametric method applicable to data in any dimension. {\cite{friedman1979multivariate} further considered the $K$-minimum spanning tree ($K$-MST)\footnote{$K$-MST: an undirected graph built as the union of the $1$st, $\cdots$ $K$th MSTs, where the 1st MST is the minimum spanning tree, and the $k$th $(k> 1)$ MST is a tree connecting all observations that minimizes the sum of distance across edges subject to the constraint that it does not contain any edges in the 1st, $\cdots$ $(k-1)$th MST(s).}, with $K=O(1)$, to achieve higher power.} Subsequent work extended OET to other similarity graphs. Notable examples include the $K$-nearest neighbor graph ($K$-NNG) \citep{schilling1986multivariate,henze1988multivariate} and the cross-match graph \citep{rosenbaum2005exact}. 

More recently, \cite{chen2017new} renovated the test statistic by incorporating a key pattern caused by the curse of dimensionality and introduced the $\emph{\text{generalized edge-count test}}$ (GET). GET on $5$-MST exhibits substantial power improvement over OET across a broad range of alternatives. Since then, additional graph-based tests have been developed. The \emph{weighted edge-count test} (WET) \citep{chen2018weighted} was designed to enhance performance under location alternatives, while the \emph{max-type edge-count test} (MET) \citep{chu2019asymptotic} exhibits performance similar to GET and has certain advantages in change-point detection scenarios. Futher theoretical advancements include the work of \cite{bhattacharya2020asymptotic}, who studied the asymptotic behavior of OET on $K$-NNG and $K$-MST with a fixed $K$ under general alternatives, using the theory of stabilizing geometric graphs in a Poissonized setting. More recently, \cite{zhu2024limiting} studied the asymptotic distributions of OET, GET and MET on graphs ranging from sparse to dense. Their findings suggest that using a dense $K$-MST or $K$-NNG, where $K$ scales with $N^\beta$ (with $N$ as the sample size and $\beta > 0$), significantly increases power under general alternatives compared to using a constant $K$.

\begin{figure}[!b]
\centering
\includegraphics[width=\textwidth]{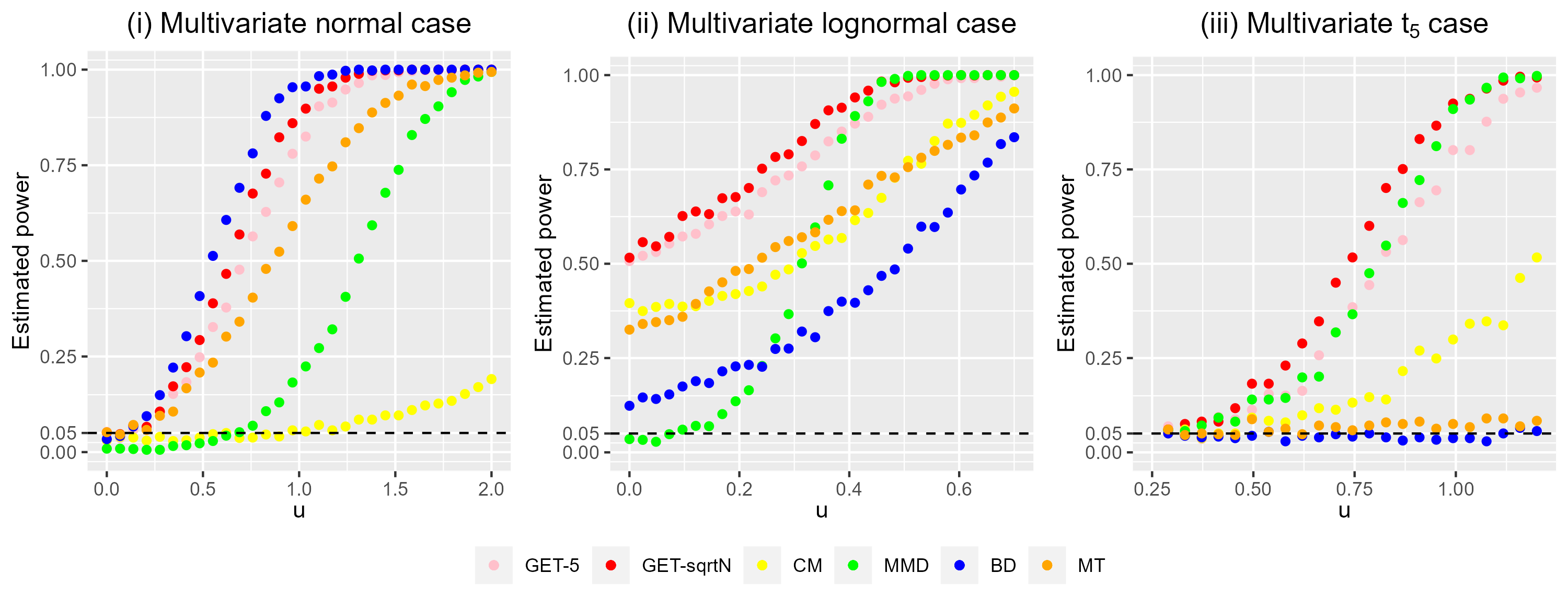}
\caption{Estimated power for different two-sample tests.}\label{estimated_power_various_tests}
\end{figure}

Other nonparametric two-sample tests have also been proposed, including those based on Maximum Mean Discrepancy (MMD) \citep{gretton2008kernel, gretton2012kernel, gretton2012optimal}, Ball Divergence \citep{pan2018ball}, and measure transportation \citep{deb2021multivariate}. Figure \ref{estimated_power_various_tests} shows the estimated power of GET on the 5-NNG (GET-5) and on the $\sqrt{N}$-NNG (GET-sqrtN), the cross match test (CM) \citep{rosenbaum2005exact}, the test based on MMD (MMD) \citep{gretton2012kernel}, the test based on the Ball Divergence (BD) \citep{pan2018ball}, and a mutivariate rank-based test (MT) \citep{deb2021multivariate} under the following scenarios: 
\begin{enumerate}
    \item[(i)] $X_1,{\cdots},X_m \iidsim N(\mathbf{0}_d, \Sigma_d(0.5))$, $Y_1,{\cdots},Y_n \iidsim N(\frac{u}{\sqrt{d}}\mathbf{1}_d,$ $ \Sigma_d(0.5)+\frac{u}{\sqrt{d}}\mathbf{I}_d)$,
    
    \item[(ii)] $X_1,{\cdots},X_m \iidsim {\text{Lognormal}}(\mathbf{0}_d,\Sigma_d(0.6))$, $Y_1,{\cdots},Y_n \iidsim {\text{Lognormal}}(\mathbf{u}_1,\Sigma_d(0.2))$,
    
    \item[(iii)] $X_1,{\cdots},X_m\iidsim t_5(\mathbf{0}_d,\Sigma_d(0.6))$,$Y_1,{\cdots},Y_n \iidsim t_5(\mathbf{u}_2,\Sigma_d(0.6))$,
\end{enumerate}
where $\mathbf{0}_d$ is a $d$-dimensional vector with elements $0$, $\mathbf{1}_d$ is a $d$-dimensional vector with elements $1$, $\mathbf{u}_1$ is a $d$-dimensional vector with first $\sqrt{d}$ elements equal to $u$ and the remaining elements equal to $0$, $\mathbf{u}_2$ is a $d$-dimensional vector with first $d^{1/3}$ elements equal to $u$ and the remaining elements equal to $0$, $\mathbf{I}_d$ is a $d$-by-$d$ identity matrix, $\Sigma_d(r) = (r^{|i-j|})_{1\leq i,\space j\leq d}$, $m = n = 100$ and $d=500$. From Figure \ref{estimated_power_various_tests}, we see that the GET test on the $K$-NNG  has good performance under location and scale differences for symmetric and asymmetric distributions, while other tests that work well under one setting could fail under some other settings.

The good performance of GET on the $K$-NNG and $K$-MST can be attributed to the test statistic's sensitivity to structure patterns in these graphs that arise from the curse of dimensionality (see \cite{chen2017new}). However, while the statistic itself has been adapted for high-dimensional regimes, the underlying $K$-NNG and $K$-MST remain prone to dimensionality-induced issues such as hub formation, which can undermine the overall effectiveness of graph-based tests (see Section \ref{sec:2} for details). To address this limitation, we propose a new and more robust graph construction that is substantially less sensitive to dimensionality effects and, when incorporated into GET, leads to consistent and often significant gains in power.

Graph-based tests are not limited to the two-sample testing problem; they can also be extended to offline change point or change interval detection \citep{chen2015graph, chen2023graph, zhang2021graph}, as well as to online change point detection in streaming data \citep{chen2019sequential, chu2022sequential}. In this paper, we focus primarily focus on the two-sample testing problem, which serves as the foundation for these broader extensions, and also provide some results for offline change point detection to illustrate the broader potential of the proposed graph.

The remainder of the paper is organized as follows. Section \ref{sec:1.2} analyzes the impact of dimensionality on the performance of GET with the $K$-NNG, with a particular focus on its influence on the graph structure. Section \ref{sec:2} introduces the proposed robust graph. Section \ref{sec:3} establishes the asymptotic properties of GET with the new graph, under both the null and alternative hypotheses. Section \ref{sec:4} presents extensive numerical comparisons with state-of-the-art methods for both two-sample testing and change-point detection. In Section \ref{sec:real_example}, we illustrate the practical utility of the robust graph in real-data applications. The paper concludes with a discussion in Section \ref{sec:conclusion}.

\section{Dimensionality effects on GET with the $K$-NNG}\label{sec:1.2}

For graph-based methods, GET on the $K$-MST or the $K$-NNG is commonly recommended due to their relatively high power across a broad range of alternatives \citep{chen2017new, chu2019asymptotic}.  For simplicity, in the following, we refer to this subset of methods as `graph-based methods' unless otherwise specified. Since using the $K$-MST or the $K$-NNG generally yield similar results, our primary focus will be on GET with the $K$-NNG, due to its relative ease of constructions. Additional results for GET on the $K$-MST are provided in the Supplemental Material.

We are concerned with identifying which types of observations or graph structures could affect the performance of these methods. We begin by examining the effect of outliers, defined as observations that are far from the majority of the data. For exploration, we use a simple setting: $X_1, {\cdots}, X_{100} \iidsim N(\mathbf{0}_d, \mathbf{I}_d)$, $Y_1, {\cdots}, Y_{100} \iidsim N(\mathbf{0}_d, \sigma^2\mathbf{I}_d)$. We consider two types of perturbations:
\begin{itemize}
    \item[ (1)] Random perturbation: mislabel the sample labels of 5 randomly chosen observations, i.e. if a randomly chosen observation belongs to sample $X$, we label it as sample $Y$, and vice versa.
    \item[ (2)] Outlier perturbation: mislabel the sample labels of 5 observations that are furthest away from the center of the data.
\end{itemize}

\begin{figure}
    \centering
    \includegraphics[scale=0.52]{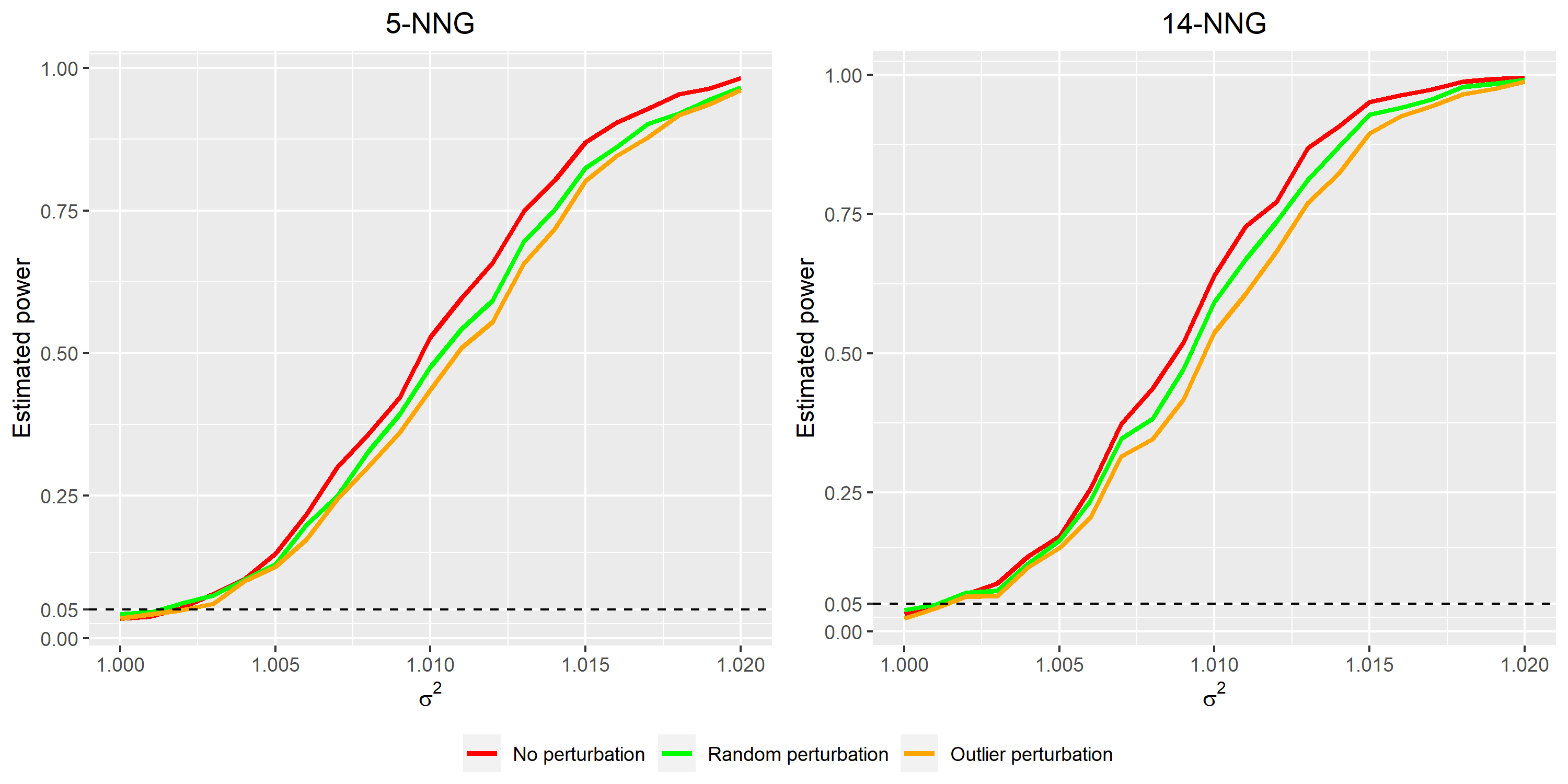}
    \caption{Estimated power of GET on the $5$-NNG and the $14$-NNG under no perturbation (red), random perturbation (green), and outlier perturbation (orange).}
    \label{Estimated power on 5NNG and 14NNG}
\end{figure}

Figure \ref{Estimated power on 5NNG and 14NNG} plots the estimated power of GET on $5$-NNG and $14$-NNG ($14\approx \sqrt{100+100}$), with $\sigma^2$  ranging evenly from 1 to 1.02 and $d = 1,000$, under no perturbation, random perturbation and outlier perturbation.
We observe that, compared to random perturbation, mislabeling observations farthest from the center decreases the power of test slightly more. This indicates that outliers are generally a bit more influential than randomly selected observations. However, the decrease in power is not substantial. This result is expected because the number of edges in the $K$-NNG that connect to the far-away observations is relatively small as other observations typically do not consider these distant observations as their nearest neighbors. Therefore, the outliers have limited influence on the test statistic.

Then, following the same reasoning, if there are observations that connect to many other observations in the similarity graph, will they significantly affect the method? To investigate this, we examine another type of perturbation:
\begin{itemize}
    \item [(3)] Hub perturbation: mislabel the sample labels of 5 observations with the highest degrees in the graph.
\end{itemize}

\begin{figure}[!]
    \centering
    \includegraphics[scale = 0.52]{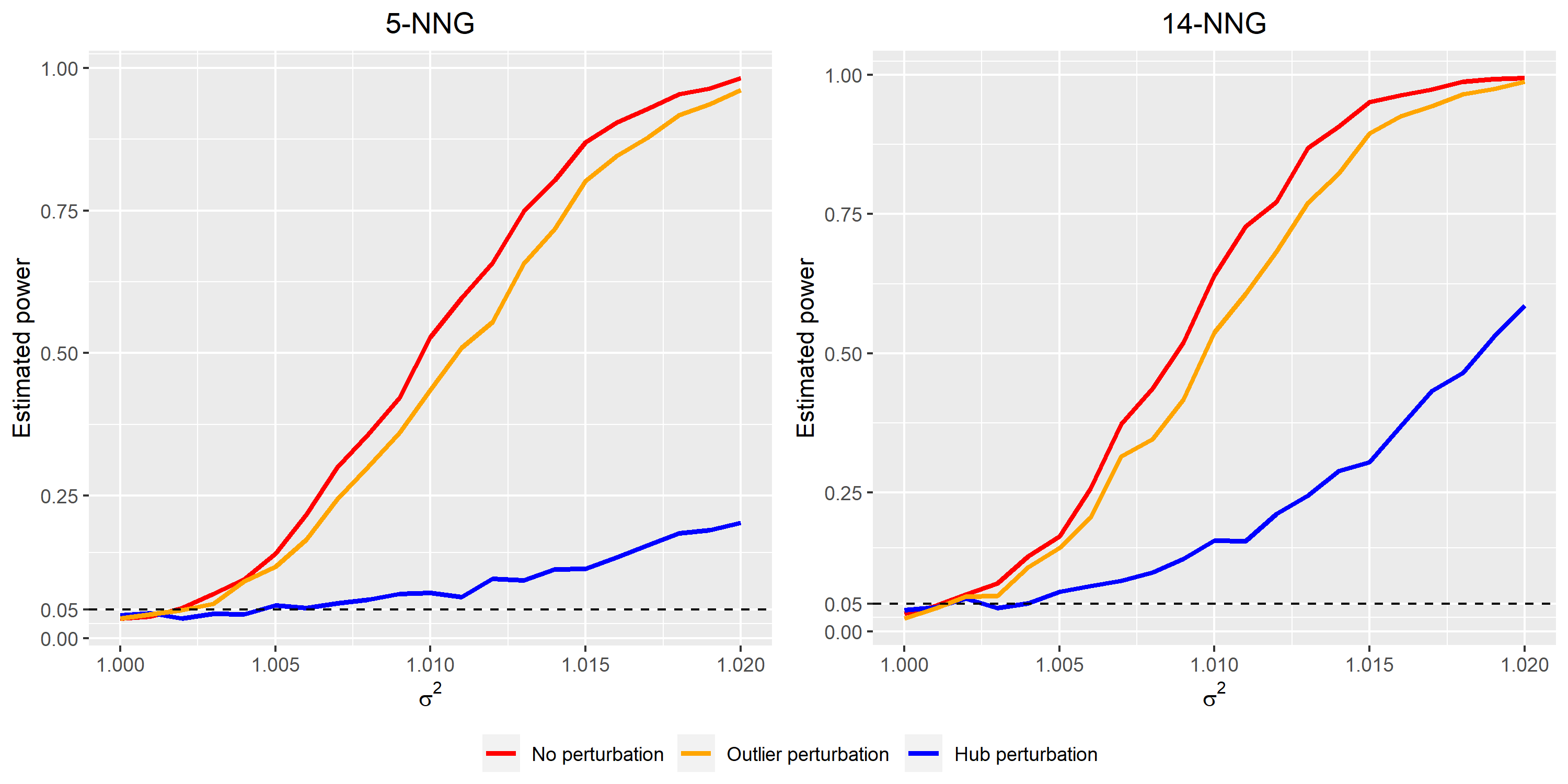}
    \caption{Estimated power of GET on the $5$-NNG and the $14$-NNG under no perturbation (red), outlier perturbation (orange), and hub perturbation (blue).}
    \label{Estimated power with inlier}
\end{figure}

Figure \ref{Estimated power with inlier} plots the estimated power of GET under the same setting as in Figure \ref{Estimated power on 5NNG and 14NNG} but with hub perturbation. We observe that hub perturbation leads to a substantial drop in test power. While using a denser graph (right panel of Figure \ref{Estimated power with inlier}) can partially mitigate this effect, there remains a significant reduction in power under hub perturbation. This sensitivity to hubs arises from the method's reliance on edge counts -- observations with high degrees can heavily influence the statistic, thereby highly affecting the test outcome. 

\begin{figure}[t]
    \centering
    \includegraphics[scale = 0.52]{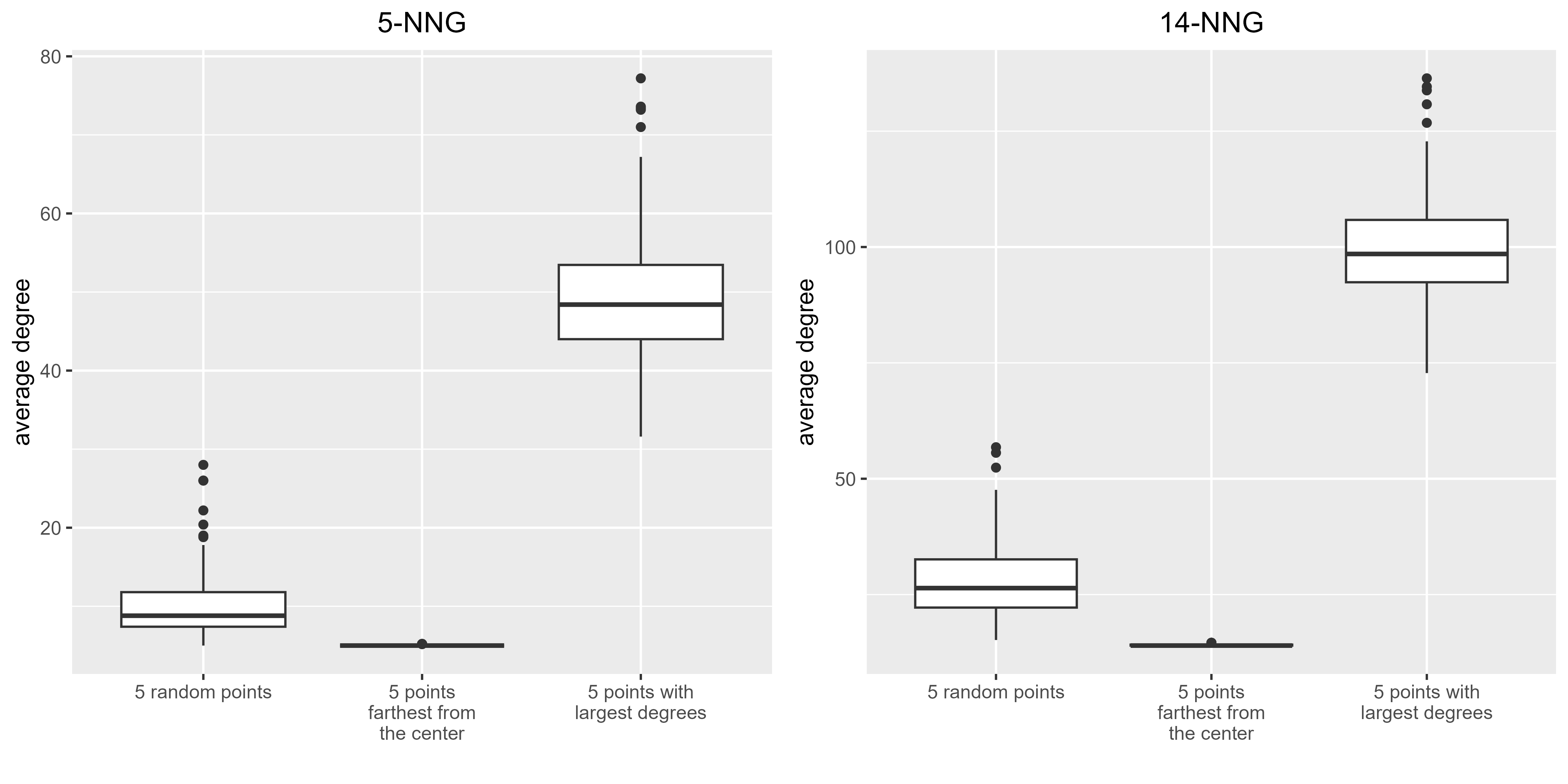}
    \caption{Average degree of perturbed observations in the $K$-NNG under the settings used in Figures \ref{Estimated power on 5NNG and 14NNG} and \ref{Estimated power with inlier}, with $\sigma^2 = 1.02$.}
    \label{degrees of mislabeled points}
\end{figure}

To evaluate how extreme the degrees of these hubs are relative to other observations, we plot the boxplots of the average degrees of the selected observations under the three types of perturbations (Figure \ref{degrees of mislabeled points}). We find that the average degree of five observations with the highest degrees in the $5$-NNG is approximately 50, compared to about 10 for five randomly selected observations. On the other hand, the average degree of the five observations farthest from the center is very small. A similar pattern is observed in the $14$-NNG.

\subsection{Relationship between hubs and dimensionality}\label{sec:hub}
The above example highlights the substantial impact of hubs on the method's performance. Notably, in moderate to high dimensions, the presence of hubs in the $K$-NNG is common. \cite{radovanovic2010hubs} investigated the phenomenon of hubness in the $K$-NNG for data drawn from a single distribution and showed that the degree distribution becomes significantly right-skewed as dimensionality increases.  We observe the same phenomenon  for data drawn from two distributions.

\begin{figure}
   \centering
    \includegraphics[width=0.98\textwidth]{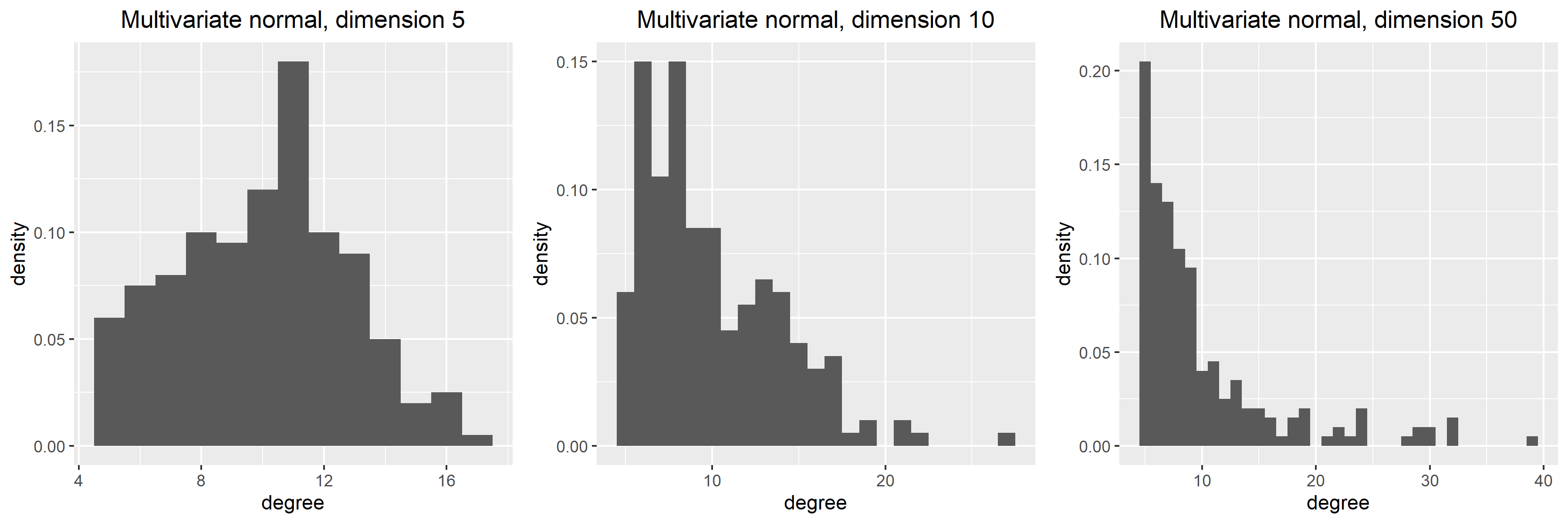}
    \includegraphics[width=0.98\textwidth]{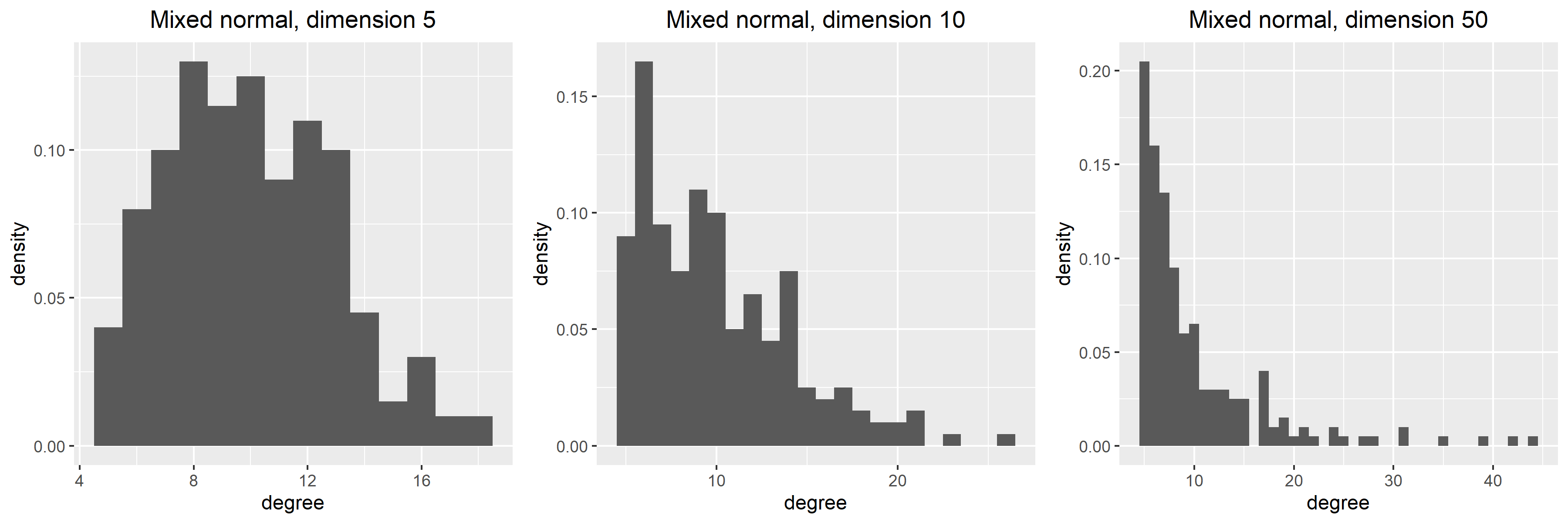}
    \caption{Empirical degree distributions of the $5$-NNG for data drawn from the standard multivariate normal distribution (top panel) and the earlier two-distribution example with $\sigma^2=1.02$ (bottom panel).}
    \label{degree_distribution_dim_5_500}
\end{figure}

Figure \ref{degree_distribution_dim_5_500} plots the empirical degree distributions of the $5$-NNG under two settings: the standard multivariate normal distribution (top panel) and the earlier two-distribution example with $\sigma^2 = 1.02$ (bottom panel), across dimensions 5, 10, and 50. In both cases, the degree distributions become right-skewed as the dimension increases.

\begin{figure}
    \centering
    \includegraphics[width=0.98\textwidth]{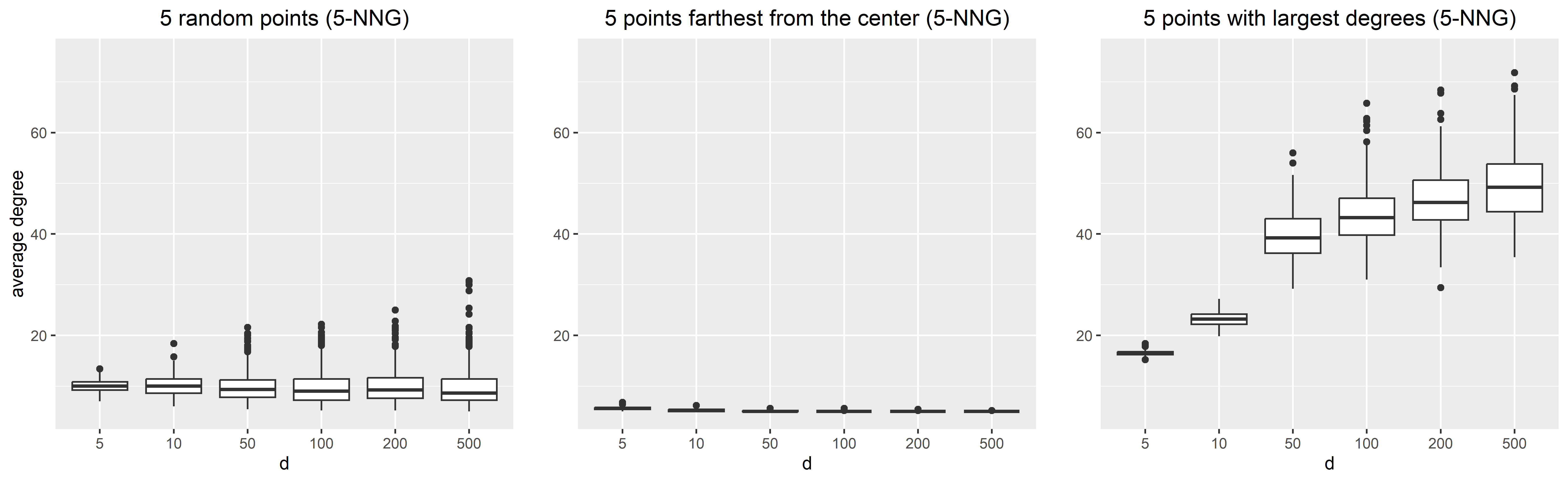}
    \includegraphics[width=0.98\textwidth]{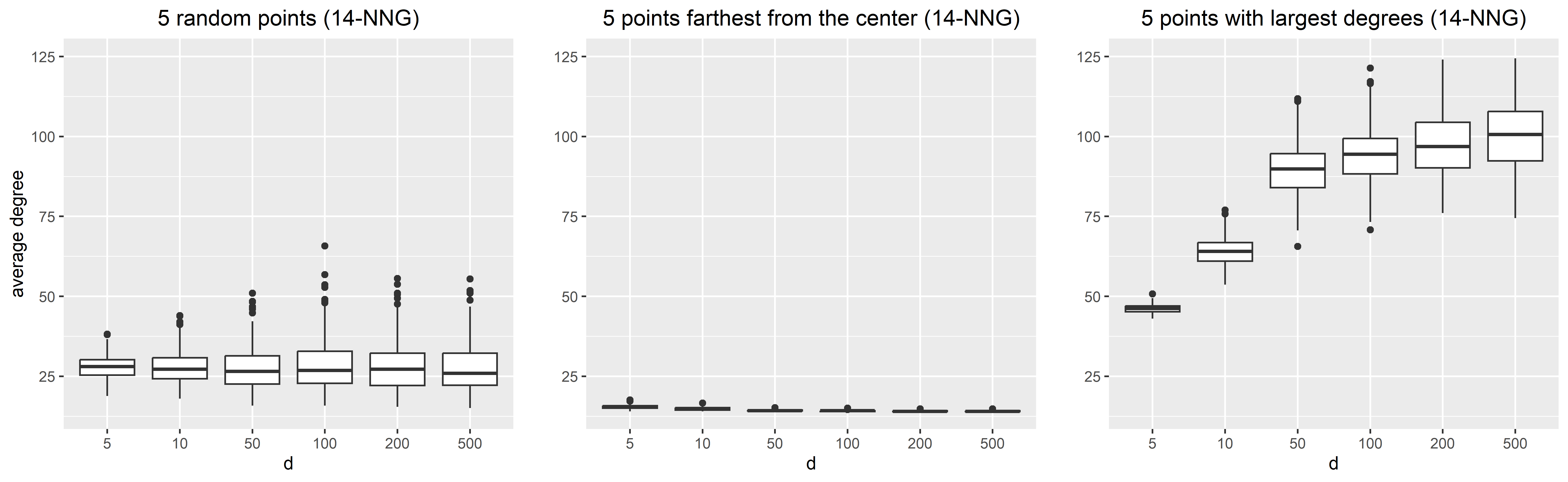}
    \caption{Boxplots of the average degree of selected points across different dimensions.}
    \label{avg_deg_vs_dim}
\end{figure}

The boxplot in Figure \ref{avg_deg_vs_dim} illustrates the average degree of  selected observations under the three types of perturbations across a range of dimensions, from $d=5$ to $d=500$. At low dimensions $(d=5)$, the average degree of the five observations with the highest degrees only slightly exceeds that of five randomly selected observations. However, as the dimension increases, the average degree of randomly selected observations remains relatively stable, while that of the five observations with the highest degrees significantly increases, leading to the formation of hubs. This increase becomes particularly evident when the dimension reaches around 50.

\begin{figure}
    \centering
     \includegraphics[scale = 0.52]{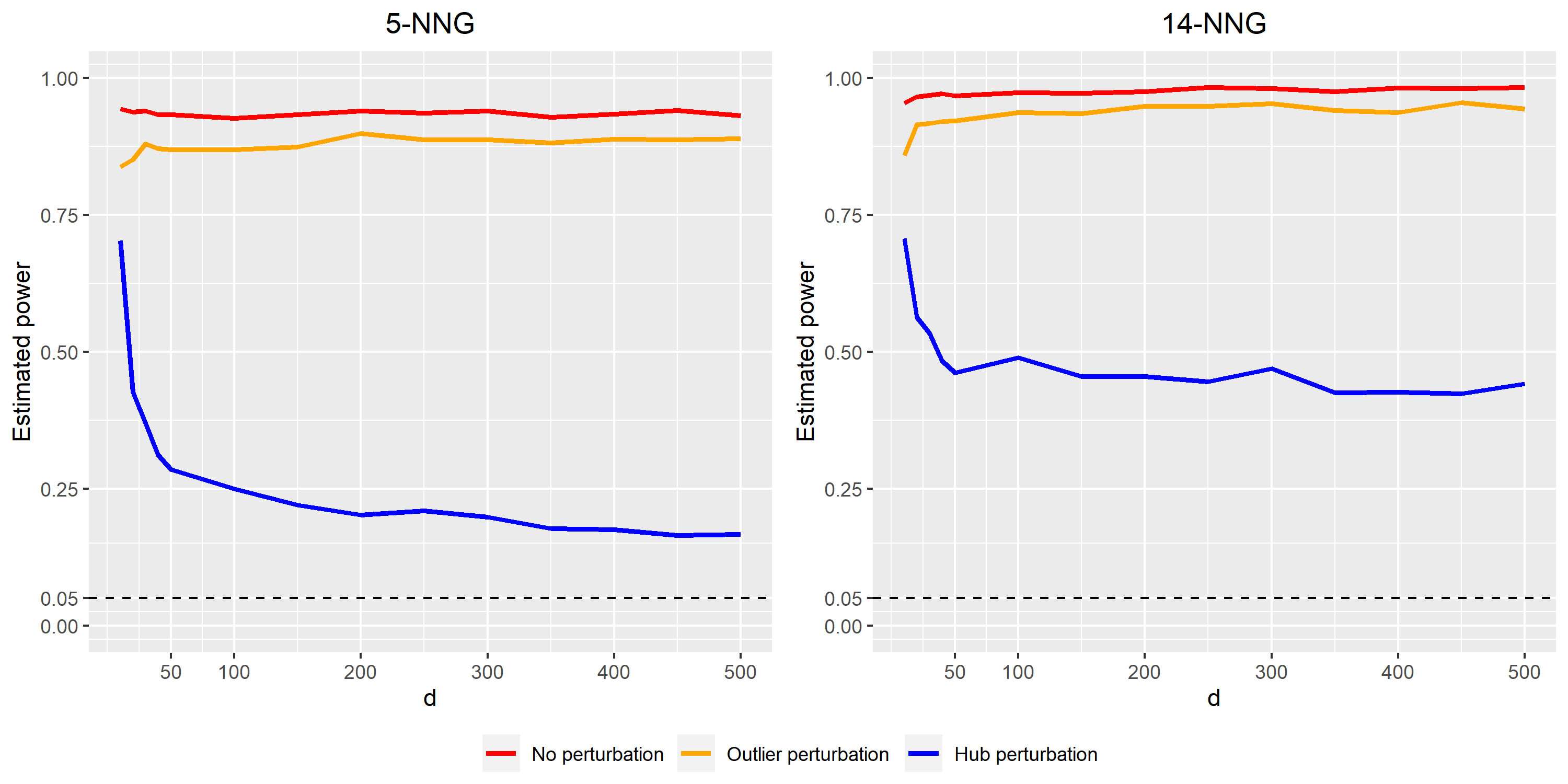}
    \caption{Estimated power of GET on the $5$-NNG and the $14$-NNG across different dimensions.}
    \label{estiamted power with d}
\end{figure}

Figure \ref{estiamted power with d} presents the estimated power of GET on both the $5$-NNG and the $14$-NNG across various dimensions. For scenarios involving no perturbation or outlier perturbation, the power remains relatively stable across all dimensions. In contrast, under hub perturbation, the estimated power starts lower than in the other two scenarios at low dimensions and shows a sharp decline until the dimension reaches around 50, after which the decline becomes more gradual. This trend aligns with the rise in the average degree of the five observations with the highest degrees, as shown in Figure \ref{avg_deg_vs_dim}.

\section{\label{sec:2} A robust graph construction to mitigate hubness}
Given that GET is sensitive to the presence of hubs, and hubs tend to form in the $K$-NNG (a similar issue also arises in the $K$-MST as well; see Supplement S4), particularly in moderate to high dimensions, we propose a modification to the $K$-NNG to limit excessively high degrees within the graph. Specifically, we introduce a penalty on the sum of squares node degrees to mitigate this issue.  

Let $Z_1,{\cdots}, Z_N$ denote the pooled observations, where $N =n+m$, and let $D(\cdot,\cdot)$ be a distance metric. Define $R_i(Z_j)$ as the rank of the distance $D(Z_i,Z_j)$ among the set $\{D(Z_i,Z_l): l \neq i\}$.  The traditional $K$-NNG seeks to minimize $\sum_{i=1}^N\sum_{x \in C_i}R_i(x)$ over all valid sets $C_i$, where each $C_i$ contains $K$ observations excluding $Z_i$. For a graph $G$, let $|G|$ denote the number of edges, and let $|G_i|$ denote the degree of the $i$-th node (counting both in-degrees and out-degrees). In the $K$-NNG, we have $|G| = KN$ and $\sum_{i=1}^N|G_i|= 2KN$. We now introduce the \emph{robust $K$-nearest neighbor graph (r$K$-NNG)}, which minimizes the following objective function over all valid sets $C_i$:
\begin{align}
\label{obj1}
    \sum_{i=1}^N\sum_{x \in C_i}R_i(x)+\lambda\sum_{i=1}^N|G_i|^2.
\end{align}
Here, $\lambda$ is a tuning parameter, with its selection discussed in Section \ref{sec: choice of lambda}. Optimizing this objective function presents a combinatorial challenge, and obtaining the global optimum is typically difficult.  As a practical solution, we propose a greedy algorithm (Algorithm \ref{alg:penalized KNN}).  Although this algorithm does not guarantee a global optimum solution, we find it to be sufficiently effective in practice.

\begin{algorithm}[!t]
\caption{Greedy algorithm for constructing the r$K$-NNG}\label{alg:penalized KNN}
Initialize $G$ with the standard $K$-NNG, compute the value of the objective function (\ref{obj1}) on $G$, and store it as $L$\;
\While{$L$ can still be further decreased}{
\begin{itemize}
    \item Randomly permute the order of the $N$ observations and label them as $1,{\cdots},N$.
    \item Let $L^{pre} \leftarrow L$
    \item \For {$i=1,\cdots, N$}{
    
    \begin{itemize}
        \item[] \hspace{-12mm} \For{$j = 1, j\leq N, j\neq i$}{

    \begin{itemize}
        \item[ ]  \hspace{-22mm} * Compute $W_{i}(j) = R_i(Z_j) + \lambda |G_j^\star|^2$, where \[
            \hspace{-40mm}|G_j^\star| = \begin{cases}
                |G_j|, & \text{if node $j$ is one of neighbors of node $i$ in $G$;} \\
                |G_j| + 1, & \text{otherwise.}
            \end{cases}
            \]
        \item[] \hspace{-22mm} *Identify the $K$ nodes with the smallest $W_{i}(j)$ values among $\{W_{i}(j) \text{ with } j\neq i\}$
        \item[] \hspace{-22mm} *Construct $G^\prime$ by replacing the $K$ outgoing edges from node $i$ in $G$ with edges to \\
        \hspace{-19mm}the $K$ nodes. Compute objective function (\ref{obj1}) on $G^\prime$, and denote the value as $L^\prime$.
        \item[] \hspace{-22mm} *If $L^\prime <L$, update $G$ with $G^\prime$ and Let $L = L^\prime$
    \end{itemize}
    }
    \end{itemize}}
    \item if $L = L^{pre}$, stop the while loop.
\end{itemize}
}
 Return the graph $G$ and $L$.
\end{algorithm}

\begin{remark}
In the objective function (\ref{obj1}), the regularization term involves the total degree $|G_i|$, which is effectively equivalent to using only the in-degree, since the out-degree for each node is fixed at $K$ in the r$K$-NNG.

\end{remark}

\begin{remark}
The objective function (\ref{obj1}) is not restricted to rank-based formulations; it can also be defined directly in terms of the distance metric $D(\cdot,\cdot)$. In this case, the robust graph construction can be obtained by solving the following optimization problem:
\begin{align*}
    &\min_{C_i\text{'s}} \sum_{i=1}^N\sum_{x\in C_i}D(Z_i,x)+\lambda\sum_{i=1}^N|G_i|^2 \text{ subject to} Z_i\notin C_i, |C_i| = K.
\end{align*}
\end{remark}

\begin{remark}\label{robust K MST}
A similar idea can be used to extend the $K$-MST to a robust version, referred to as the robust $K$-MST (r$K$-MST). Let $R(Z_i,Z_j)$ denote the rank of the distance $D(Z_i,Z_j)$ among all pairwise distances.
The r$K$-MST is a $K$-spanning tree that minimizes the following objective function 
\begin{align*}
    \sum_{(z_i,z_j) \in T}R(z_i,z_j)+\lambda\sum_{i=1}^N |T_i|^2,
\end{align*}
where $T$ is a $K$-spanning tree, and $|T_i|$ is the degree of node $i$ in $T$.
\end{remark}

\subsection{Performance of GET on the robust $K$-NNG (r$K$-NNG)} \label{sec: mitigated effect of K-RNNG}
\begin{figure}[!t]
    \centering
    \includegraphics[scale = 0.52]{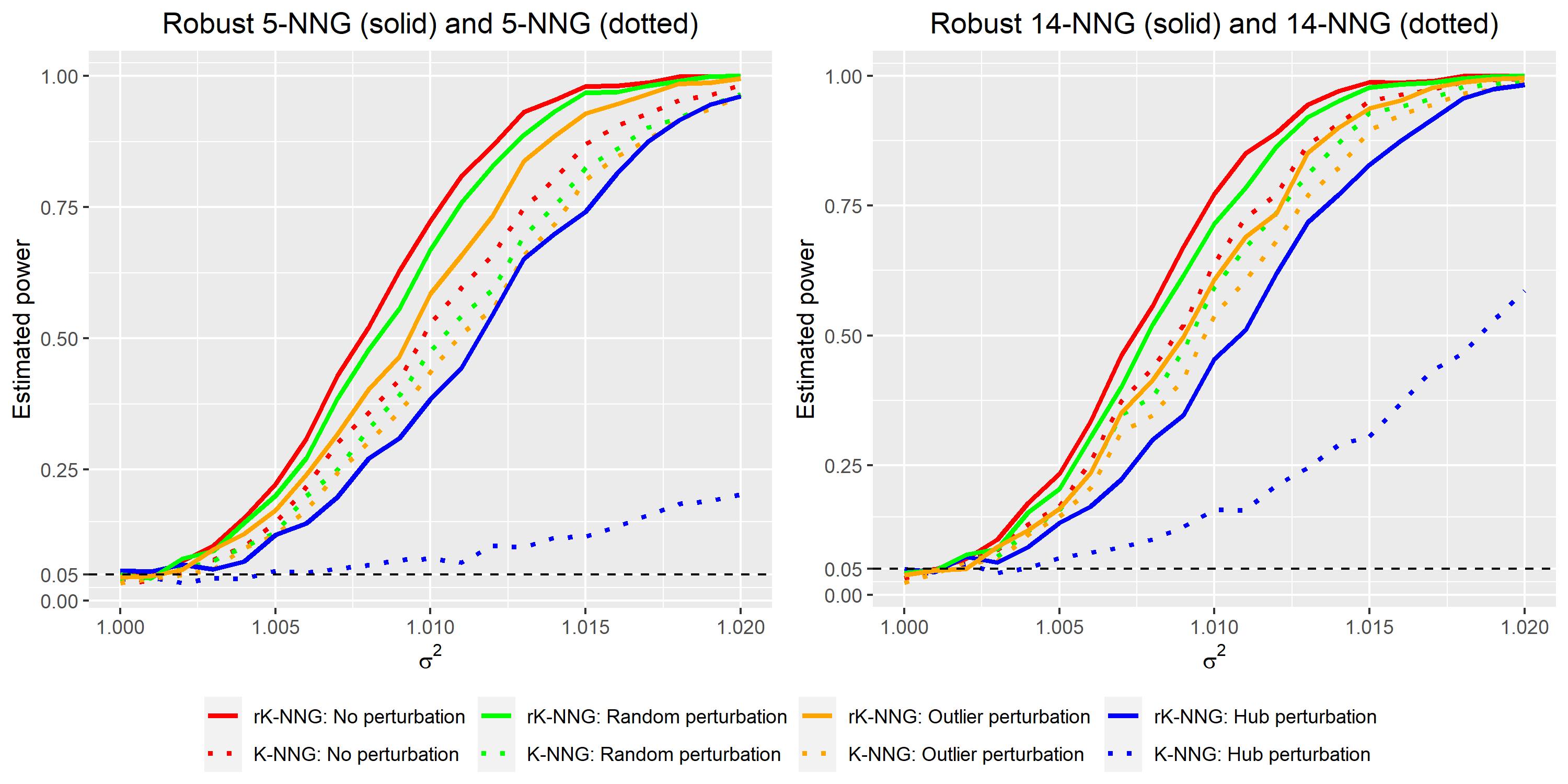}
    \caption{Estimated power of GET with the robust $K$-NNG (solid) compared to the standard $K$-NNG (dotted) across scenarios with no perturbation and with different types of perturbations.}
    \label{estimated_power on K-RNNG}
\end{figure}

\begin{figure}[!t]
    \centering
    \includegraphics[scale = 0.52]{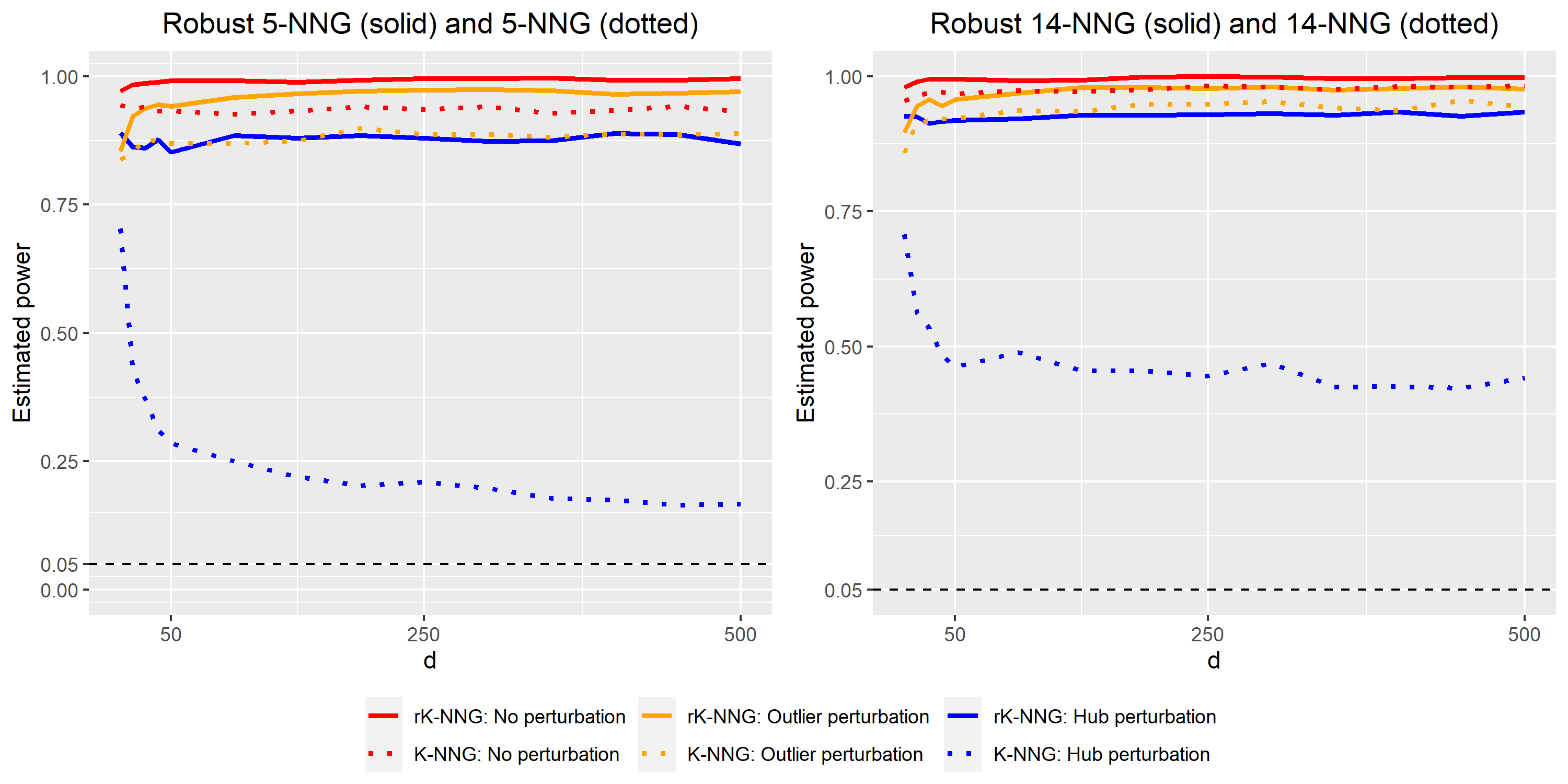}
    \caption{Estimated power of GET with robust vs. standard $K$-NNG across different dimensions.}
    \label{power of KRNNG with d}
\end{figure}

We now examine the performance of GET when applied to the robust $K$-nearest neighbor graph (r$K$-NNG).
 Figures \ref{estimated_power on K-RNNG} and \ref{power of KRNNG with d} present results from the same simulation settings used in Figures \ref{Estimated power on 5NNG and 14NNG}, \ref{Estimated power with inlier}. and \ref{estiamted power with d}. Across all settings, GET on the robust $K$-NNG shows substantially higher power under hub perturbation, as seen from the clear gap between the blue solid lines (robust $K$-NNG) and the blue dotted lines (standard $K$-NNG).

\begin{figure}[!t]
    \centering
    \includegraphics[width = 0.98\textwidth]{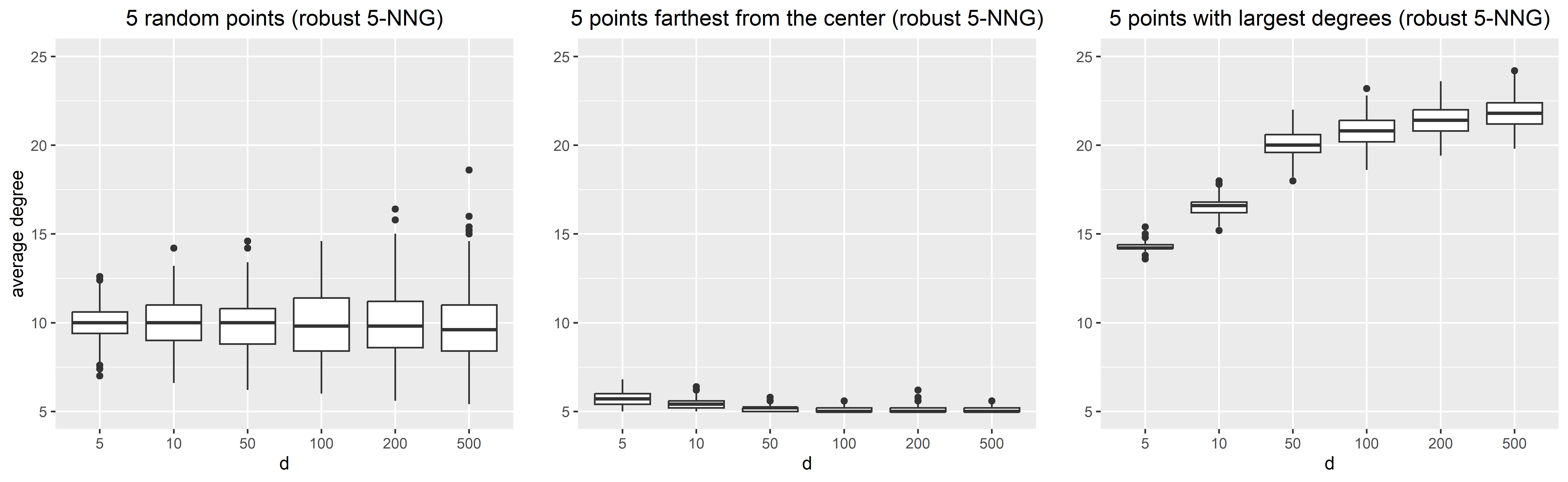}
    \includegraphics[width = 0.98\textwidth]{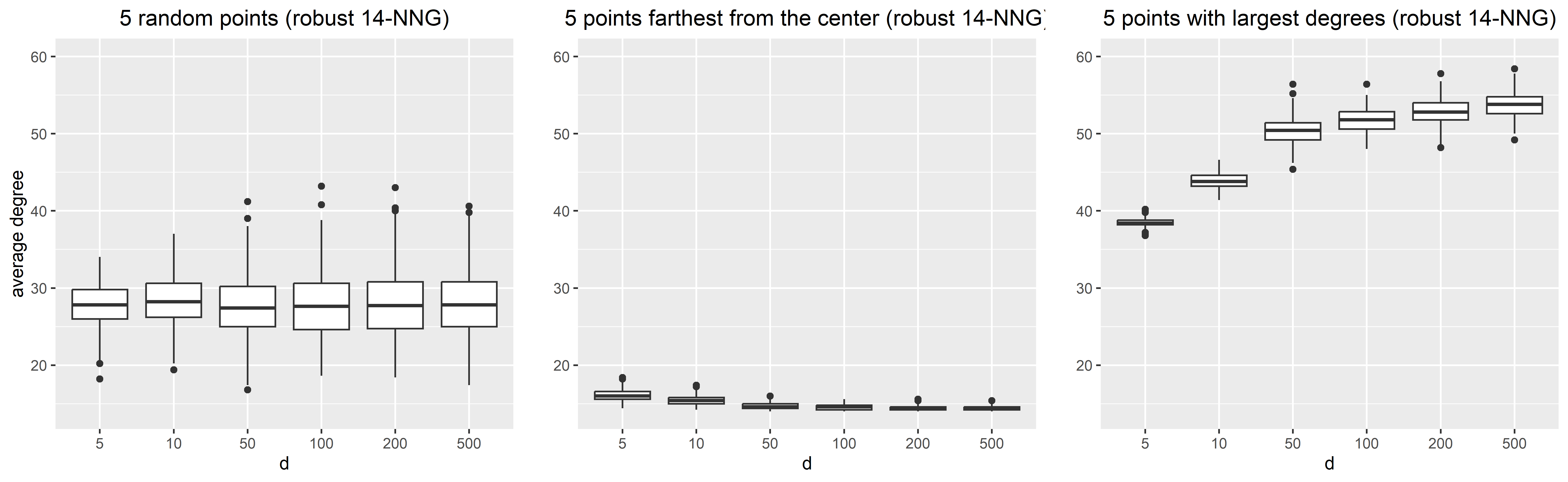}
    \caption{Boxplots of the average degree of selected points in the robust $5$-NNG and robust $14$-NNG across different dimensions.}
    \label{avg_deg_vs_dim_of K-RNNG}
\end{figure}

Figure \ref{avg_deg_vs_dim_of K-RNNG}  further illustrates the effect of hub mitigation. It shows the average degree of perturbed observations in the robust $5$-NNG and $14$-NNG under the same setting to that in Figure \ref{avg_deg_vs_dim}. As dimension increases from 5 to 500, the average degree of the top-5 highest-degree nodes grows dramatically in the standard $5$-NNG (from 17 to 50), while in the robust $5$-NNG the increase is much more modest (from 14 to about 22). This confirms that the robust graph construction effectively suppresses hub formation as dimensionality grows.

\begin{figure}[!t]
    \centering
    \includegraphics[width = \textwidth]{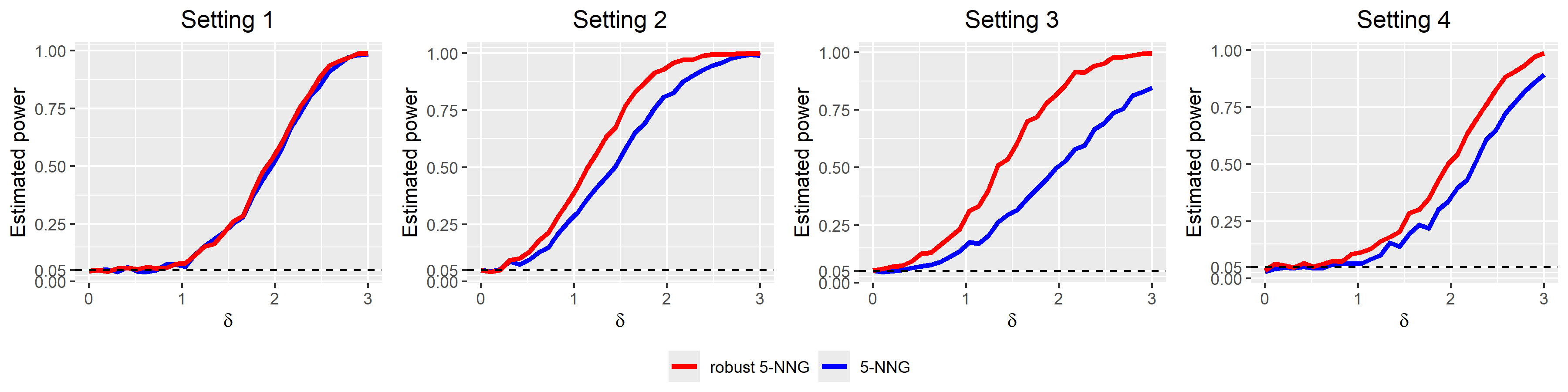}
    \caption{Estimated power of GET with robust (red) vs. standard (blue) $5$-NNG across multiple distributions.}
    \label{KRNNG_vs_KNNG}
\end{figure}

Beyond hub perturbation, the robust $K$-NNG also improves power in the presence of outlier and random perturbations, and even in settings without perturbation (Figures \ref{estimated_power on K-RNNG} and \ref{power of KRNNG with d}). To assess robustness across a wider range of distributions, we further consider multivariate normal, lognormal, and $t$ distributions with both mean and variance differences (Settings 1–4 below). The results, summarized in Figure \ref{KRNNG_vs_KNNG}, show that GET on the robust $K$-NNG performs comparably to the standard $K$-NNG when only mean differences are present, but offers a clear advantage when variance differences are involved.

\begin{itemize}
    \item[1.] $X_1, {\cdots}, X_m \iidsim N(\mathbf{0}_d, \Sigma_d(0.5))$, $Y_1,{\cdots},Y_n \iidsim N(\frac{\delta}{\sqrt{d}}\mathbf{1}_d, $ $\Sigma_d(0.5))$,
    \item[2.] $X_1,{\cdots}, X_m \iidsim N(\mathbf{0}_d, \Sigma_d(0.5))$, $Y_1,{\cdots},Y_n \iidsim N(\frac{\delta}{\sqrt{d}}\mathbf{1}_d, $ $\Sigma_d(0.5)+\frac{\delta}{\sqrt{d}}\mathbf{I}_d)$,
    \item[3.] $X_1,{\cdots}, X_m \iidsim \text{Lognormal}(\mathbf{0}_d, \Sigma_d(0.5))$, $Y_1,{\cdots},Y_n \iidsim \text{Lognormal}(\frac{\delta}{\sqrt{d}}\mathbf{1}_d, \Sigma_d(0.5))$,
    \item[4.] $X_1,{\cdots}, X_m \iidsim \text{Multivariate t}_5(\mathbf{0}_d, \Sigma_d(0.5))$, $Y_1,{\cdots},Y_n \iidsim \text{Multivariate t}_5(\frac{\delta}{\sqrt{d}}\mathbf{1}_d,\\ \Sigma_d(0.5)+\frac{\delta}{\sqrt{d}}\mathbf{I}_d)$,
    
\end{itemize}

This behavior can be explained by the structure of the GET statistic:
\begin{align*}
    S = \begin{pmatrix}R_1-\epp(R_1),& R_2-\epp(R_2) \end{pmatrix}
    \left(\varp\binom{R_1}{R_2}\right)^{-1} \begin{pmatrix}R_1-\epp(R_1) \\R_2-\epp(R_2)\end{pmatrix},
\end{align*}
where  $R_1$ is the number of edges in the graph connecting observations within sample $X$, and $R_2$ is the number of edges connecting observations within sample $Y$. Here, $\epp$, $\varp$ and $\covp$ are the expectation, variance and covariance under the permutation null distribution, which assigns equal probability to each selection of $m$ observations from  $N$  as sample $X$. 
 \cite{chu2019asymptotic} showed that $S$ can be decomposed into two components:
 $S = (Z_w^\P)^2 + (Z_\d^\P)^2,$
where $$Z_w^\P = \frac{R_w-\epp(R_w)}{\varp(R_w)^{1/2}}, \quad Z_{\d}^\P = \frac{R_{\d}-\epp(R_\d)}{\varp(R_\d)^{1/2}},\quad \covp(Z_w^\P, Z_\d^\P) = 0,$$ with $R_w = R_1(n-1)/(N-2)+R_2(m-1)/(N-2)$ and $R_\d= R_1-R_2$. Prior research has shown that $Z_w^\P$ is primarily sensitive to mean differences, and $Z_\d^\P$ is more effective for detecting variance differences \citep{chu2019asymptotic, song2022asymptotic, liu2022fast}.  From \cite{chu2019asymptotic}, we have that 
\begin{align*}
    &\epp(R_\d) = \frac{m-n}{N}|G|, \varp(R_\d) = \frac{mn(m-1)(n-1)}{N(N-1)(N-2)(N-3)}\left(\frac{m-2}{n-1}+\frac{n-2}{m-1}+2\right)V_G,
\end{align*}
where $V_G = \sum_{i=1}^N(|G_i|-2|G|/N)^2$.  In the robust $K$-NNG, the penalty on node degrees reduces the variability term $V_G$ in the expression for $\varp(R_\d)$. Consequently, the variance of $R_\d$ is reduced, which enhances the power of $Z_\d^\P$ and makes the robust $K$-NNG especially effective in scenarios involving scale or variance differences.

\subsection{Choice of $\lambda$}\label{sec: choice of lambda}
We next investigate the choice of the tuning parameter $\lambda$, which controls the degree of hub penalization in the robust $K$-NNG. To this end, we consider a range of scenarios involving symmetric, asymmetric, and heavy-tailed distributions:
\begin{itemize}
    \item[(i)] $X_1,{\cdots},X_m \iidsim N(\mathbf{0}_d,\Sigma_d(0.5))$, $Y_1,{\cdots},Y_n \iidsim N(\mathbf{0}_d,\delta\Sigma_d(0.5))$ with $m = 200$, $n =100$, $d= 500$ and $\delta = 1.03$;
    \item[(ii)] $X_1,{\cdots},X_m \iidsim \text{lognormal}(\mathbf{0}_d,\Sigma_d(0.5))$, $Y_1,{\cdots},Y_n \iidsim \text{lognormal}(\delta\mathbf{1}_d,\Sigma_d(0.5))$ with $m = 100$, $n =200$, $d= 1000$ and $\delta = 0.05$;
    \item[(iii)] $X_1,{\cdots},X_m \iidsim \text{lognormal}(\mathbf{0}_d,\Sigma_d(0.5))$, $Y_1,{\cdots},Y_n \iidsim \text{lognormal}(\mathbf{0}_d,\delta\Sigma_d(0.5))$ with $m = 100$, $n =100$, $d= 100$ and $\delta = 1.15$;
    \item[(iv)] $X_1,{\cdots},X_m \iidsim \text{t}_5(\mathbf{0}_d,\Sigma_d(0.5))$, $Y_1,{\cdots},Y_n \iidsim \text{t}_5(\mathbf{0}_d,\delta\Sigma_d(0.5))$ with $m = 100$, $n =100$, $d= 500$ and $\delta = 1.35$;
    \item[(v)] $X_1,{\cdots},X_m \iidsim \text{t}_5(\mathbf{0}_d,\Sigma_d(0.5))$, $Y_1,{\cdots},Y_n \iidsim \text{t}_5(\delta\mathbf{1}_d, \Sigma_d(0.5))$ with $m = 300$, $n =100$, $d= 500$ and $\delta = 0.095$.
\end{itemize}
The values of $\delta$ are chosen so that the power of GET is between 0.4 and 0.6 when $\lambda=0$.

\begin{figure}
    \centering
    \includegraphics[scale = 0.7]{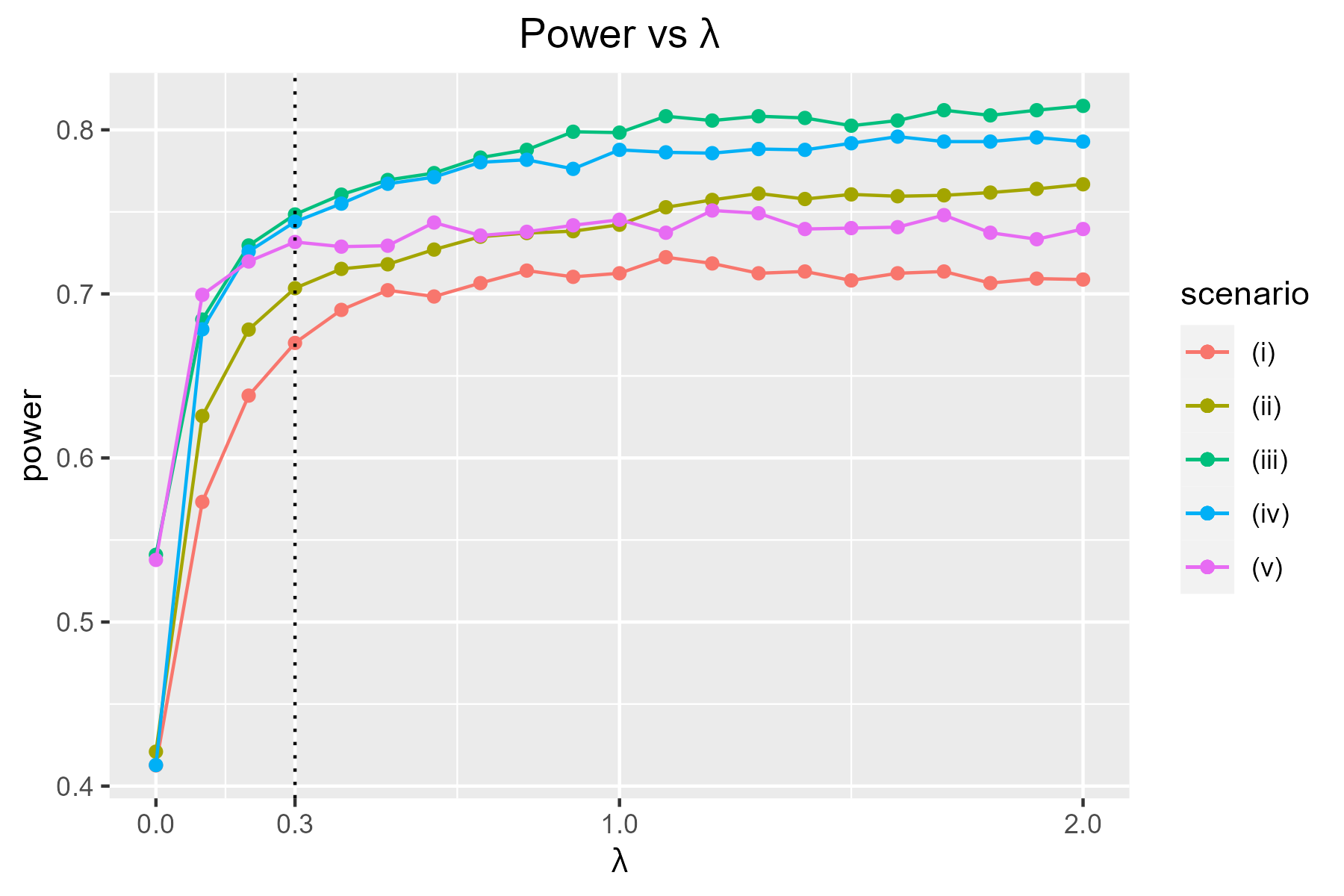}
    \caption{Estimated power across different values of $\lambda$ under various scenarios.}
    \label{lambda_selection}
\end{figure}

Figure \ref{lambda_selection} shows the effect of $\lambda$ on power.  Across all scenarios, the power increases sharply as $\lambda$ increases from 0 to 0.3. Beyond 0.3,  the gain in power becomes more gradual, suggesting that $\lambda=0.3$ provides a reasonable default choice. 

\begin{figure}
    \centering
    \includegraphics[scale = 0.7]{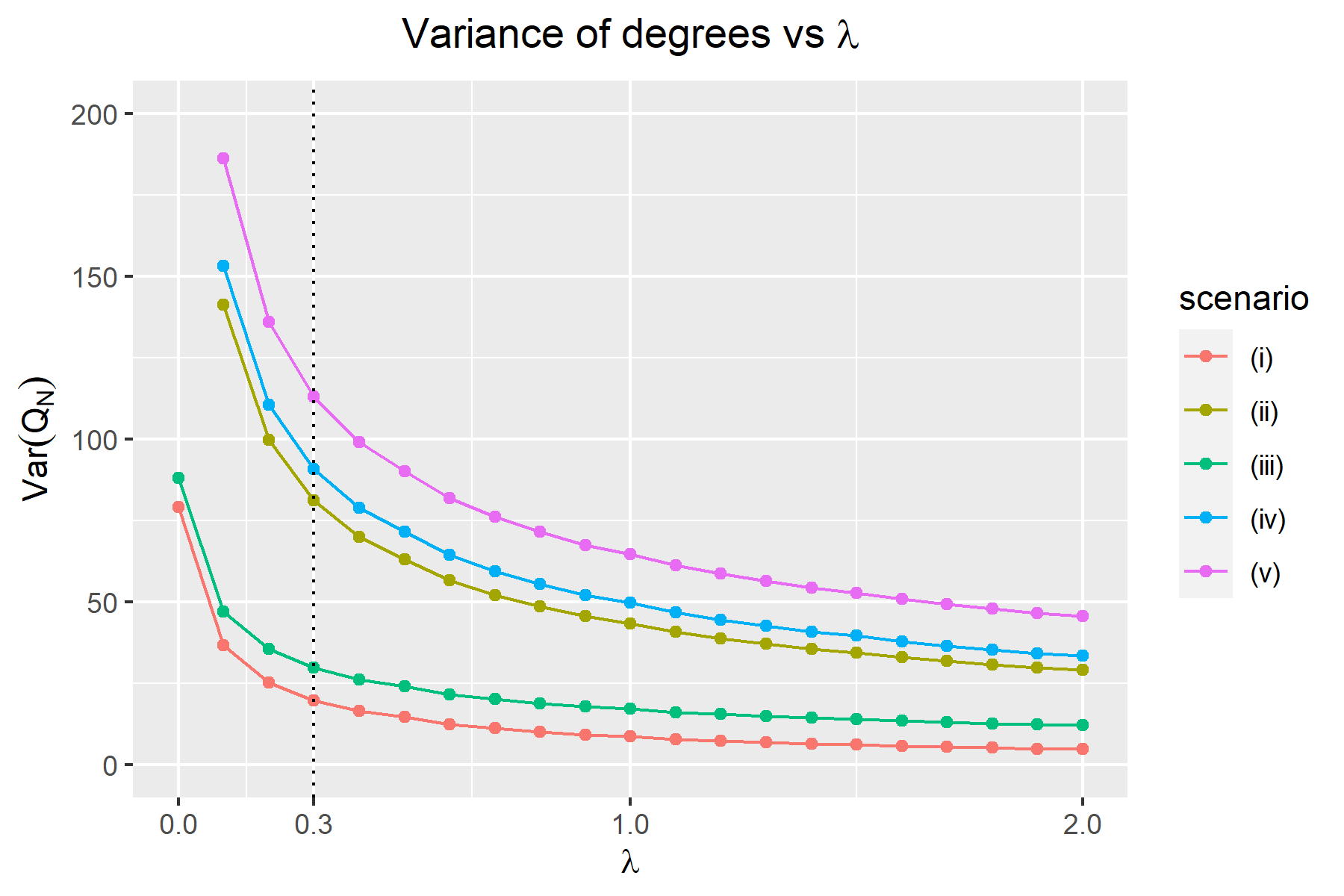}
    \caption{$\var(Q_N)$ across different values of $\lambda$ under various scenario.}
    \label{lambda_variance}
\end{figure}

Based on the analysis in Section \ref{sec: mitigated effect of K-RNNG}, the quantity $\sum_{i=1}^N|G_i|^2$ plays a key role in controlling the variance of $R_\d$. Since this quantity depends on $N$,  we instead plot $\sum_{i=1}^N(|G_i|-2K)^2/N = (\sum_{i=1}^N|G_i|^2-4K^2)/N$, which can be viewed as the variance of node degrees and is denoted by $\var(Q_N)$ (where $Q_N$ is defined in Theorem \ref{th2} in Section \ref{sec:4}). Figure \ref{lambda_variance} shows $\var(Q_N)$ with respect to $\lambda$ in these scenarios. Consistent with the power results, $\var(Q_N)$ decreases rapidly up to $\lambda=0.3$ and then declines more gradually. This reinforces the choice of $\lambda=0.3$ as a robust default. 

In practice, direct estimation of power is not feasible. However, $\var(Q_N)$ can be readily computed for candidate values of $\lambda$. Thus, practitioners may use the $\var(Q_N)$ curve to guide the selection of $\lambda$ in real applications, while $\lambda=0.3$ serves as a broadly effective default.

\section{Asymptotic properties of the GET statistic on the r$K$-NNG} \label{sec:3}

\cite{zhu2024limiting} derived  the best-known sufficient conditions  for the validity of the asymptotic distribution of the GET statistic on undirected graphs. In this section, we extend their results to directed graphs and demonstrate that the robust $K$-NNG graphs satisfies these conditions with an appropriate choice of $\lambda$. Before presenting the results, we first introduce some essential notation.

Given a directed graph $G$, let the ordered pair $(i,j)$ represent a directed edge pointing from node $i$ to node $j$. For each node $i$, we define $G_i$ as the set of edges that involve node $i$, $G_{i,2}$ as the set of edges that share at least one node with an edge in $G_i$, $node_{G_i}$ as the set of nodes  connected in $G_i$ excluding  node $i$, and $node_{G_{i,2}}$ as the set of nodes  connected in $G_{i,2}$ also excluding  node $i$. Let $\Tilde{d}_i$ denote the centered degree of the $i$-th node, defined as $\Tilde{d}_i = |G_i| - 2|G|/N$, and let $V_G = \sum_{i=1}^N \Tilde{d}_i^2$ represent the degree variation. We further define $N_0$ as the number of edges whose reversed edge is also in $G$, i.e. $N_0 = \sum_{\left(i,j\right)\in G}1_{\{(j,i)\in G\}}$, and $N_{sq}$ to be the number of combinations of four edges that form a quadrilateral, such as $(i,j),(j,s),(s,t),(t,i)$.

Additionally, we use $\xrightarrow{\mathcal{D}}$ to denote convergence in distribution, and use `the usual limit regime' to refer to $N\rightarrow \infty$ and $\lim_{N\rightarrow \infty}\frac{m}{N} = p\in (0,1)$. In the following, $a_n= o(b_n)$ or $a_n\prec b_n$ means that $a_n$ is dominated by $b_n$ asymptotically, i.e. $\lim_{n\rightarrow \infty} \frac{a_n}{b_n}=0$, $a_n\precsim b_n$ or $a_n = O(b_n)$ means that $a_n$ is asymptotically bounded above by $b_n$ (up to a constant factor), and $a_n= \Theta(b_n)$ or $a_n \asymp b_n$ means that $a_n$ is asymptotically bounded both above and below by $b_n$ (up to constant factors). We use $a\wedge b$ to denote $\min\{a,b\}$. For two sets $S_1$ and $S_2$, $S_1\backslash S_2$ denotes the set  containing elements in $S_1$ but not in $S_2$.

\cite{zhu2024limiting} established sufficient conditions for the asymptotic distribution of the test statistic on undirected graphs using the `locSCB’ approach.  This approach relies on the equivalence between the permutation null distribution and the conditional Bootstrap null distribution. The Bootstrap null distribution assigns each observation to either sample $X$ or sample $Y$ independently, with probabilities $\frac{m}{N}$ and $\frac{n}{N}$, respectively. Conditioning on the number of observations assigned to sample $X$ being $m$, the Bootstrap null distribution becomes the permutation null distribution. The authors applied the Stein's method that considers the first neighbor dependency (as reflected in the `loc' component of the `locSCB' approach) under the Bootstrap null distribution to derive asymptotic multivariate normality. 

We  adopt a similar approach to derive sufficient conditions for  directed graphs. A key challenge in this setting is the possibility of two directed edges between two nodes, necessitating careful consideration of certain graph-related quantities. {The sufficient conditions for general directed graph are provided in Lemma \ref{th1} with the proofs in Supplement \textcolor{blue}{S2.1}.} For the robust $K$-NNG, a more concise result can be obtained, as stated in Theorem \ref{th2}, with the corresponding proof given in Supplement S1.

\begin{lemma}\label{th1}
For a directed graph $G$ with $|G| = O(N^\alpha), 1\leq \alpha<2$, under conditions 
\begin{align*}
   &\sum_{i=1}^{N}\left|G_{i}\right|^{2}=o\left(|G|^{\frac{3}{2}}\right), \sum_{i=1}^N \left|\Tilde{d_i}\right|^3 = o(V_G^\frac{3}{2}), \sum_{i=1}^N \Tilde{d_i}^3 = o(V_G\sqrt{|G|}),\\
   &\sum_{i=1}^N\sum_{(i,j) \text{ or }(j,i) \in G_i}^{(i,k) \text{ or }(k,i) \in G_i, j\neq k}\Tilde{d_j}\Tilde{d_k} = o(|G|V_G), \quad N_{sq} = o(|G|^2).
\end{align*}
 in the usual limit regime, we have $S\xrightarrow{\mathcal{D}} \chi_2^2$ under the permutation null distribution.
\end{lemma}

\begin{theorem}\label{th2}
Let $Q_N$ be the random variable generated from the degree distribution of the robust $K$-NNG with $N$ nodes. Assume $K = \Theta(1)$. If $\lambda$ is chosen such that $\var(Q_N)>0$ and $ \max\{Q_N\}\precsim N^{\frac{1}{2} -\beta}$ for some $\beta > 0$, then $S\xrightarrow{\mathcal{D}} \chi_2^2$ under the permutation null distribution in the usual limit regime.

\end{theorem}

\begin{remark}
Theorem \ref{th2} requires that the variance of degree distribution of the robust $K$-NNG is asymptotically bounded away from zero when choosing $\lambda$.  This ensures that the GET statistic is well defined -- when $\var(Q_N)=0$, the degrees of all nodes are the same, and $\varp\binom{R_1}{R_2}$ becomes singular. This situation arises when an extremely large $\lambda$ is used. Additionally, large values of $\lambda$ diminish the utilization of the similarity information contained in the first term of the objective function \eqref{obj1}, making such choices of $\lambda$ less desirable. Therefore, we tend not to choose a very large $\lambda$ in practice.
\end{remark}

\begin{theorem}[Consistency under fixed dimensions] \label{th3}
For two samples generated 
from continuous multivariate distributions in a fixed-dimensional Euclidean space, if the graph is the r$K$-NNG with $K = \Theta(1)$ and $\lambda \geq 0$, then GET is consistent against all alternatives in the usual limiting regime.
\end{theorem}

Proof for Theorems \ref{th3} is provided in Supplement Material S1.  For consistency in high dimensions, we adopt the assumptions from \cite{biswas2014distribution}. They provided three assumptions to ensure that the weak law of large numbers is applicable to $||X_1-X_2||_2^2/d$, $||Y_1-Y_2||_2^2/d$ and $||X_1-Y_1||_2^2/d$, which ensures that Assumption 1 is valid.

\begin{assumption}
    Let $X_1$, $X_2$ be independent copies of $X$, and $Y_1$, $Y_2$ be independent copies of $Y$. There exists $\sigma_1^2$, $\sigma_2^2 >0$ and $v^2$ such that $||X_1-X_2||_2^2/d$, $||Y_1-Y_2||_2^2/d$ and $||X_1-Y_1||_2^2/d$ converge to $2\sigma_1^2$, $2\sigma_2^2$ and $\sigma_1^2+\sigma_2^2+v^2$ in probability, respectively, when $d\rightarrow \infty$.
\end{assumption}

\begin{theorem}[Consistency under high dimensions]\label{th4}
Assume that the distributions $F_X$ and $F_Y$ satisfy Assumption 1. 
Then, for GET on the r$K$-NNG with $0<\lambda < (\sqrt{8NK+4N-8K}-\sqrt{8NK})^2/16$ and $\min\{m,n\} > K+2\lambda+\sqrt{8\lambda KN}$, we have $$\lim_{d\rightarrow \infty}P(S>\chi_2^2(1-\alpha)) = 1,$$ for any fixed $\alpha \in (0,1)$, provided that one of the following conditions holds:
\begin{itemize}
    \item [(1)] $|\sigma_1^2 - \sigma_2^2| < v^2$, and $N > 2.5 + \frac{\xi}{K}+\sqrt{0.25+3\frac{\xi}{K} + \frac{\xi^2}{K^2}}$,
    \item [(2)]
    $\sigma_1^2 - \sigma_2^2 > v^2$, and $N>\frac{n^2\xi^2}{2m^2K^2}\left(\sqrt{\frac{K}{\lambda}}+\sqrt{\frac{K}{\lambda}+\frac{2mK}{n\xi}(1+\frac{K}{2\lambda}+\frac{mK}{n\xi}-K)}\right)^2$, 
    \item[(3)] $\sigma_2^2 - \sigma_1^2 > v^2$, and $N>\frac{m^2\xi^2}{2n^2K^2}\left(\sqrt{\frac{K}{\lambda}}+\sqrt{\frac{K}{\lambda}+\frac{2nK}{m\xi}(1+\frac{K}{2\lambda}+\frac{nK}{m\xi}-K)}\right)^2$. 
\end{itemize}
Here, $\xi =\chi_2^2(1-\alpha)$ denotes the $(1-\alpha)$-quantile of the $\chi_2^2$ distribution.
\end{theorem}

Proof for Theorems \ref{th4} is provided in Supplement Material S1. 
In practice, since the exact relationship among the parameters $\sigma_1^2$, $\sigma_2^2$ and $v^2$ is typically unknown, the largest of the three sample size requirements may be used as a conservative estimate of the minimum $N$. For instance, with $\alpha = 0.05$, $\lambda = 0.3$, $m=n$, and $K = 5$, the conservative threshold is $N \geq 69$. If $m/n = 2$ or $n/m = 2$ with the same values of $\alpha, \lambda$ and $K$,  the conservative threshold becomes $N \geq 214$.

\section{Numerical studies}\label{sec:4} In this section, we evaluate the performance of GET on the r$K$-NNG by comparing it with several state-of-the-art methods in both  two-sample testing and change-point detection problems.

\subsection{Two-sample testing}
We consider GET on the robust 5-NNG (New), 5-MST, and $\sqrt{N}$-MST, alongside several popular tests: the cross-match test \citep{rosenbaum2005exact} (CM), the Ball divergence test \citep{pan2018ball} (BD), the mutivariate rank-based test \citep{deb2021multivariate} (MT), the Adaptable Regularized Hotelling's $\text{T}^2$ test \citep{li2020adaptable} (ARHT), the kernel test based on minimum mean discrepancy \citep{gretton2012kernel} (MMD), and the generalized kernel two-sample test \citep{song2023generalized} (gKernel). These methods are compared under simulation settings that encompass both symmetric and asymmetric cases, as well as Gaussian-tailed and heavy-tailed distributions.
\begin{itemize}
    \item[TS1.] $X_1,{\cdots},X_m \iidsim N(\mathbf{0}_d, \Sigma_d(0.5))$, $Y_1,{\cdots},Y_n \iidsim N(\frac{\delta}{\sqrt{d}}\mathbf{1}_d,$ $\Sigma_d(0.5)+\frac{\delta}{\sqrt{d}}\mathbf{I}_d)$,
    \item[TS2.] $X_1,{\cdots},X_m \iidsim \text{lognormal}(\mathbf{0}_d,\Sigma_d(0.6))$, $Y_1,{\cdots},Y_n \iidsim \text{lognormal}(\mathbf{\Delta},\Sigma_d(0.2))$,
    \item[TS3.] $X_1,{\cdots},X_m \iidsim \text{Multivariate t}_2(\mathbf{0}_d, \Sigma_d(0.5))$, $Y_1,{\cdots},Y_n \iidsim \text{Multivariate t}_2(\frac{\delta}{\sqrt{d}}\mathbf{1}_d, \Sigma_d(0.5)+\delta\mathbf{I}_d)$,
    \item[TS4.] $X_1,{\cdots},X_m \iidsim \text{Multivariate t}_1(\mathbf{0}_d, \Sigma_d(0.5))$, $Y_1,{\cdots},Y_n \iidsim \text{Multivariate t}_1(\frac{\delta}{\sqrt{d}}\mathbf{1}_d, \Sigma_d(0.5)+\frac{\delta}{2}\mathbf{I}_d)$,
\end{itemize}
where $\mathbf{\Delta}$ a $d$-dimensional vector with the first $\sqrt{d}$ elements equal to $\delta$ and the remaining elements equal to $0$.

\begin{figure}
    \centering
    \includegraphics[scale = 0.5]{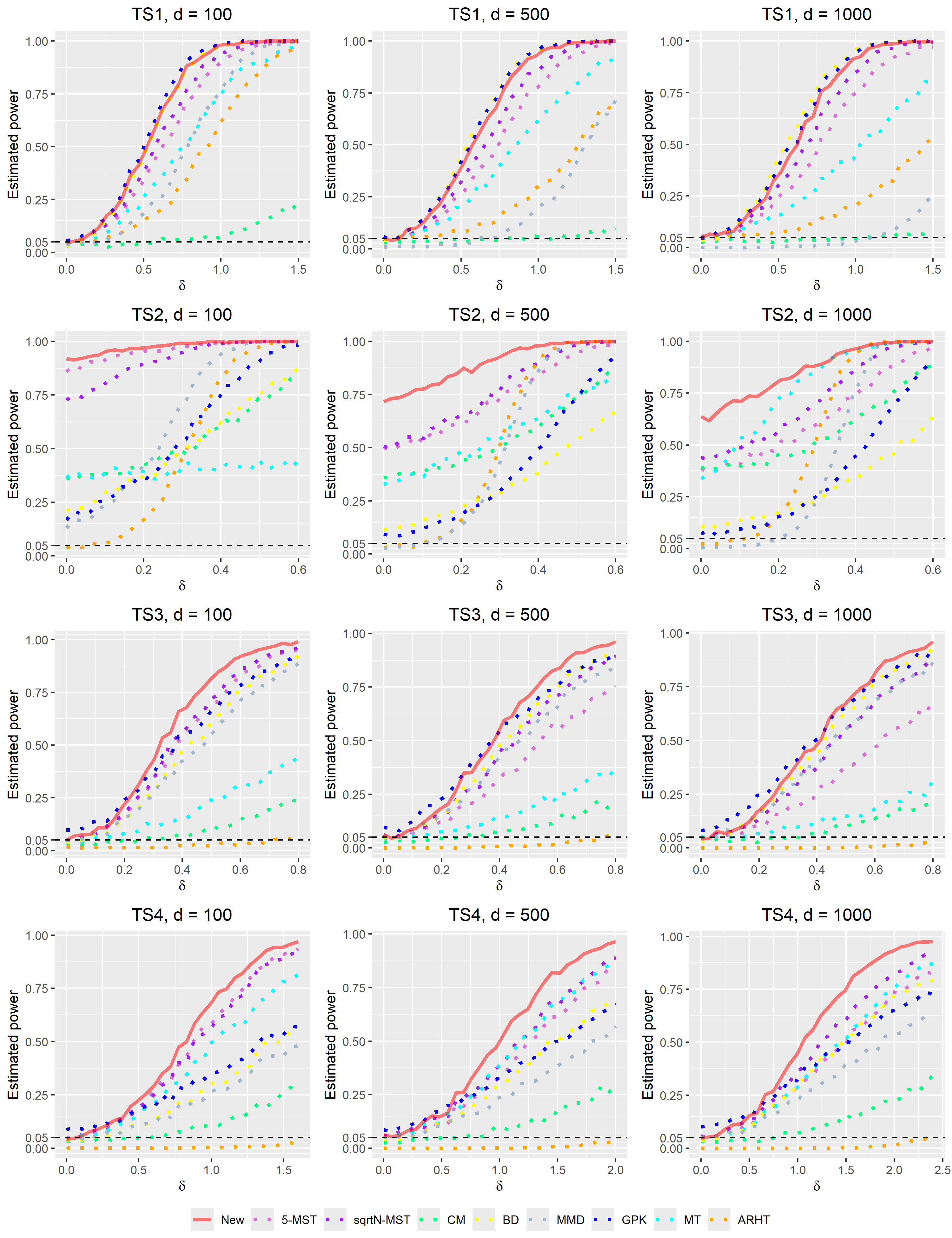}
    \caption{Estimated power of two-sample tests under four scenarios (TS1–TS4) with symmetric/asymmetric and Gaussian-/heavy-tailed distributions.}
    \label{two_sample_sim}
\end{figure}

In each scenario, we set $m=n=100$ and $d = 50, 500, 1000$. The estimated powers computed from 1000 repetitions are plotted in Figure \ref{two_sample_sim}. First, we observe that the empirical sizes of GET on the robust $5$-NNG are well controlled across all scenarios (at $\delta=0$ in Scenarios 1, 3 and 4). In Scenario 1, although the power of the new test is slightly lower than that of BD and gKernel, which exhibit the highest power in this setting, it clearly outperform them in Scenarios 2, 3 and 4, showing particularly notable improvements in Scenarios 2 and 4. Moreover, the new test consistently outperforms GET on the standard $5$-MST and GET on the standard $\sqrt{N}$-MST across all scenarios. These results highlight the advantages of using the robust $K$-NNG in practice.

\subsection{Change-point detection} \label{numeric study for change point detection}
For graph-based change-point detection, MET is often recommended over GET \citep{chu2019asymptotic,liu2022fast,song2022asymptotic}. In this section, we evaluate the performance of both MET on the robust $5$-NNG and GET on the robust $5$-NNG. We also include in the comparison the GET scan statistic on the $5$-MST, the MET scan statistic on the $5$-NNG, and the distance-based approach in \cite{matteson2014nonparametric, james2015ecp} (e.divisive).  We
consider the following simulation settings, which covers symmetric and asymmetric cases, as well as Gaussian-tailed and heavy-tailed distributions:

\begin{itemize}
    \item[CP1.]  $X_1,{\cdots},X_\tau \iidsim  N(\mathbf{0}_d, \Sigma_d(0.5))$, $X_{\tau + 1},{\cdots}, X_{N} \iidsim N(\frac{\delta}{\sqrt{d}} \mathbf{1}_d,$ $\Sigma_d(0.5)+\frac{\delta}{\sqrt{d}}{\mathbf{I}_d})$;
    
    \item[CP2.] $X_1,{\cdots},X_\tau \iidsim N(\mathbf{0}_d, \mathbf{I}_d)$, $X_{\tau + 1},{\cdots}, X_{N} \iidsim N(\mathbf{0}_d, \Sigma_d(\delta))$;
    
    \item[CP3.] $X_1,{\cdots},X_\tau \iidsim \text{Multivariate t}_5(\mathbf{0}_d, \Sigma_d(0.5))$, $X_{\tau + 1},{\cdots}, $ $X_{N}\iidsim \text{Multivariate t}_5(\frac{\delta}{d}\mathbf{1}_d,\\ \delta\mathbf{I}_d + \Sigma_d(0.5))$;
    \item[CP4.] $X_1,{\cdots}, X_\tau \iidsim \text{lognormal}(\mathbf{0}_d, \Sigma_d(0.5))$, $X_{\tau + 1}, {\cdots}, X_N \iidsim \text{lognormal}(\frac{\delta}{d}\mathbf{1}_d, \Sigma_d(0.5)+\frac{\delta}{\sqrt{d}}\mathbf{I}_d)$.
\end{itemize}

In each setting, we set $N= 400$, the true change-point $\tau$ at $100,200$ or $300$, and the dimension $d$ to be either $100$ or $500$. The estimated power is calculated as the proportion of trials (out of 1000) that reject the null hypothesis of no change-point at 0.05 significance level, and accuracy is defined as the proportion of trials with a significant $p$-value and an estimated change-point $\hat \tau$ satisfying $|\hat \tau-\tau|\leq 10$. The estimated power and accuracy for Setting 1 are plotted in Figure \ref{cp_normal}, and the results for the other settings are provided in Appendix \ref{plots of change-point study}. We observe that the MET and GET scan statistics based on the robust $5$-NNG exhibit strong power and accuracy in all settings, while others methods may perform well in some settings but poorly in others.

\begin{figure}
    \centering
    \includegraphics[width=\textwidth]{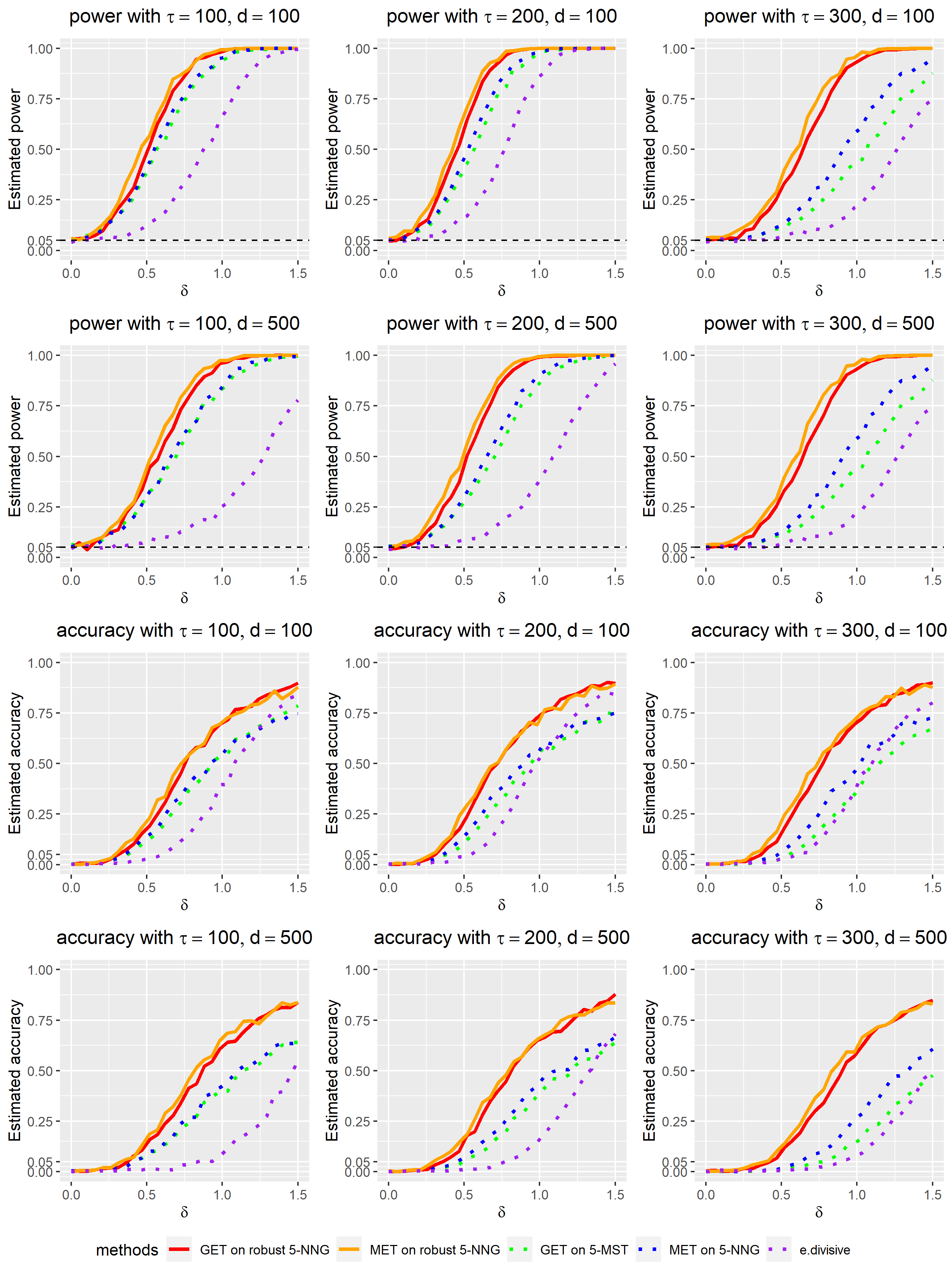}
    \caption{Estimated power and detection accuracy of change-point methods for Setting CP1.}
    \label{cp_normal}
\end{figure}

\section{Real-data examples}\label{sec:real_example}

We illustrate the practical advantages of the proposed robust graph through three real-data applications: (i) comparing correlation matrices of brain regions between patients and controls, (ii) identifying gene expression differences across T cell subtypes, and (iii) detecting changes in New York City taxi travel patterns. Across all examples, incorporating the robust $K$-NNG leads to more reliable detection of distributional differences compared to its standard counterpart and competing methods.

\subsection{Correlation matrix of brain regions}

We first analyze correlation matrices from brain imaging data assembled by \cite{xu2023data}, which compiled MRI scans from six sources -- ABIDE, ADNI, PPMI, Matai, TaoWu, and Neurocon -- covering conditions such as autism, Alzheimer’s disease, and Parkinson’s disease. Brain regions of interest were defined under several parcellation strategies, including AAL \citep{tzourio2002automated}, HarvardOxford (HO) \citep{makris2006decreased}, Schaefer \citep{schaefer2018local}, k-means Clustering \citep{abraham2014machine}, and  Ward Clustering \citep{abraham2014machine}, and correlation matrices were computed for each subject.

We focus on two datasets: \textbf{TaoWu} (20 Parkinson’s patients and 20 controls) and \textbf{ABIDE} (487 Alzheimer’s patients and 537 controls). We applied GET on the robust 5-NNG (GET-r5NNG), GET on the standard 5-NNG (GET-5NNG), Ball Divergence (BD), and a multivariate rank-based test (MT). The $p$-values are summarized in Table \ref{tab:MRI data}.

The MT method did not detect significant differences between patients and controls in either the \textbf{TaoWu} or \textbf{ABIDE} datasets under any parcellation. In contrast, the GET-5NNG identified significant differences in the \textbf{ABIDE} dataset with the AAL, HarvardOxford, and Schaefer parcellations but found no significant differences in the \textbf{TaoWu} dataset. Both GET-r5NNG and BD detected significant differences in both datasets under certain parcellation strategies. Notably, in the \textbf{TaoWu} dataset with the Schaefer parcellation, GET-r5NNG rejected the null while BD returned only a marginally small but non-significant $p$-value. Across all cases where both GET-r5NNG and BD detected differences, GET-r5NNG consistently yielded smaller  $p$-values, demonstrating its greater sensitivity.

To better understand the improvements of the robust 5-NNG over the standard 5-NNG, we examine the values of $R_w$ and $R_\d$ for both methods, summarized in Table \ref{Rw_Rd_fMRI}. We focus on the \textbf{TaoWu} dataset (with Harvard–Oxford and Schaefer parcellations) and the ABIDE dataset (with Ward clustering), where GET-r5NNG demonstrates markedly better performance than GET-5NNG. In the TaoWu dataset, the robust 5-NNG increases the mean of $R_w$ while maintaining its variance, and substantially reduces the variance of $R_\d$. This yields higher standardized values for both $R_w$ and $R_\d$, thereby improving test power. In the \textbf{ABIDE} dataset, by contrast, the robust 5-NNG slightly reduces the standardized $R_w$ but significantly increases the standardized $R_\d$ by lowering its variance. In this case, the improvement in $R_\d$ is the primary driver of the enhanced performance.

These findings also inform the choice of parcellation. In the \textbf{TaoWu} dataset, the Harvard–Oxford parcellation appeared most informative, while in the \textbf{ABIDE} dataset, AAL, HarvardOxford, Schaefer, and Ward Clustering are all informative.

\begin{table}[]
\centering
\caption{$p$-values from different two-sample tests for patient–control comparisons of brain correlation matrices under various parcellation strategies}
\begin{tabular}{|l|l|llll|}
\hline
Data  & Parcellation             & GET-5NNG          & GET-r5NNG         & BD            & MT                \\ \hline
\textbf{TaoWu} & AAL                      & 0.83262           & 0.084         & 0.07          & 0.6234\\
\textbf{TaoWu} & HarvardOxford            & 0.05601           & \textbf{0.00117} & \textbf{0.02} & 0.6494  \\
\textbf{TaoWu} & schaefer                 & 0.49376           & \textbf{0.043}  & 0.08          & 0.8132  \\
\textbf{TaoWu} & $k$-means Clustering     & 0.19425           & 0.198           & 0.8           & 0.3177  \\
\textbf{TaoWu} & Ward Clustering          & 0.5703            & 0.385            & 0.36          & 0.3896   \\
\textbf{ABIDE} & AAL                      & \textbf{2.22E-16} & \textbf{2.07E-14} & \textbf{0.04} & 0.2837  \\
\textbf{ABIDE} & HarvardOxford            & \textbf{8.67E-09} & \textbf{1.43E-08} & \textbf{0.01} & 0.3956  \\
\textbf{ABIDE} & schaefer & \textbf{6.44E-15} & \textbf{6.66E-15} & \textbf{0.01} & 0.09191 \\
\textbf{ABIDE} & $k$-means Clustering     & 0.36483           & 0.1978           & 0.15          & 0.06893 \\
\textbf{ABIDE} & Ward Clustering          & 0.19197           & \textbf{1.61E-07} & \textbf{0.01} & 0.554  \\ \hline
\end{tabular}
\label{tab:MRI data}
\end{table}

\begin{table}[!b]
\caption{$R_w$ and $R_\d$ values and standardized statistics for GET with standard vs. robust 5-NNG in brain correlation data.}
\centering
\begin{tabular}{|l|ll|ll|ll|}
\hline
                                  & \multicolumn{2}{l|}{TaoWu - HarvardOxford} & \multicolumn{2}{l|}{Taowu - schaefer} & \multicolumn{2}{l|}{ABIDE - Ward Clustering} \\ \hline
                                  & \multicolumn{1}{l|}{5-NNG}     & r5-NNG    & \multicolumn{1}{l|}{5-NNG}  & r5-NNG  & \multicolumn{1}{l|}{5-NNG}       & r5-NNG     \\ \hline
$R_w$                             & \multicolumn{1}{l|}{55.5}      & 61        & \multicolumn{1}{l|}{50.5}   & 56      & \multicolumn{1}{l|}{1267}        & 1264      \\
$\epp(R_w)$                          & \multicolumn{1}{l|}{48.7}      & 48.7      & \multicolumn{1}{l|}{48.7}   & 48.7    & \multicolumn{1}{l|}{1277}        & 1277      \\
$\varp(R_w)$                        & \multicolumn{1}{l|}{14.1}      & 17.7      & \multicolumn{1}{l|}{11.8}   & 16.5    & \multicolumn{1}{l|}{70}          & 321       \\
$\frac{R_w-\epp(R_w)}{\sqrt{\varp(R_w)}}$    & \multicolumn{1}{l|}{1.81}      & 2.91      & \multicolumn{1}{l|}{0.52}   & 1.79    & \multicolumn{1}{l|}{-1.16}        & -0.70      \\ \hline
$R_\d$                            & \multicolumn{1}{l|}{25}        & 16        & \multicolumn{1}{l|}{21}     & 14      & \multicolumn{1}{l|}{1159}        & 858       \\
$\epp(R_\d)$                         & \multicolumn{1}{l|}{0}         & 0         & \multicolumn{1}{l|}{0}      & 0       & \multicolumn{1}{l|}{-240}        & -240      \\
$\varp(R_\d)$                       & \multicolumn{1}{l|}{251.8}     & 51.3      & \multicolumn{1}{l|}{386}    & 64      & \multicolumn{1}{l|}{1013216}     & 39513     \\
$\frac{R_\d-\epp(R_\d)}{\sqrt{\varp(R_\d)}}$ & \multicolumn{1}{l|}{1.57}      & 2.23      & \multicolumn{1}{l|}{1.07}   & 1.75    & \multicolumn{1}{l|}{1.39}        & 5.54      \\ \hline
\end{tabular}
\label{Rw_Rd_fMRI}
\end{table}

\subsection{Gene expression in T cells subtypes}

Next, we analyze scRNA-seq data from \cite{zheng2017landscape}, which profiled T cells in liver cancer. Three subtypes were studied: CD8+ cytotoxic T cells, CD4+ naive T cells, and CD4+ regulatory T cells (T\textsubscript{reg}). We compared gene expression distributions in the DNA Replication Pathway (hsa03030) from the KEGG database, a key pathway for cell proliferation and tumor progression.

The results (Table \ref{GET-Tcells}) show that, at 0.01 significance level, GET-5NNG and BD missed differences between CD8+ cytotoxic vs. CD4+ naive and between CD8+ cytotoxic vs. CD4+ T\textsubscript{reg}, while GET-r5NNG detected highly significant differences. BD confirmed only one significant comparison, and MT lacked power in all cases.

Detailed analysis of $R_w$ and $R_\d$ (Table \ref{Rw_Rd_spk}) shows that the robust graph consistently reduced the variance of $R_\d$, yielding much higher standardized values. In some comparisons, it also improved standardized $R_w$. These results highlight how the robust graph enhances the sensitivity of GET in complex, high-dimensional biological data.

\begin{table}[h]
\centering
\caption{$p$-values for pairwise comparisons of T cell subtypes}
\begin{tabular}{|c|c|c|c|c|}
\hline
T cell type                   & GET-5NNG & GET-r5NNG &BD & MT\\ \hline
CD8+ Cytotoxic vs CD4+ Naive  & 0.026    & $\mathbf{9.4*10^{-5}}$ & 0.013  &0.81         \\
CD8+ Cytotoxic vs CD4+ T\_reg & 0.025    & \textbf{0.0054} & 0.011  &0.53      \\
CD4+ Naive vs CD4+ T\_reg     & $\mathbf{6.1*10^{-10}}$   & $\mathbf{6*10^{-12}}$ & \textbf{0.01} &0.06         \\ \hline
\end{tabular}
\label{GET-Tcells}
\end{table}

\begin{table}[h]
\centering
\caption{$R_w$ and $R_\d$ values and standardized statistics for GET with standard vs. robust 5-NNG in T cell data}
\footnotesize
\begin{tabular}{|l|ll|ll|ll|}
\hline
                                  & \multicolumn{2}{l|}{CD8+ Cytotoxic vs CD4+ Naive} & \multicolumn{2}{l|}{CD8+ Cytotoxic vs CD4+ T reg} & \multicolumn{2}{l|}{CD4+ Naive vs CD4+ T reg} \\ \hline
                                  & \multicolumn{1}{l|}{5-NNG}         & r5-NNG       & \multicolumn{1}{l|}{5-NNG}         & r5-NNG       & \multicolumn{1}{l|}{5-NNG}       & r5NNG      \\ \hline
$R_w$                             & \multicolumn{1}{l|}{1033.5}        & 1056         & \multicolumn{1}{l|}{647.8}         & 655.3        & \multicolumn{1}{l|}{682.5}       & 680      \\
$\epp(R_w)$                          & \multicolumn{1}{l|}{1021.2}        & 1021.2       & \multicolumn{1}{l|}{622.1}         & 622.1        & \multicolumn{1}{l|}{619.3}       & 619.3      \\
$\varp(R_w)$                        & \multicolumn{1}{l|}{315}           & 345          & \multicolumn{1}{l|}{161}           & 175          & \multicolumn{1}{l|}{162}         & 178     \\
$\frac{R_w-\epp(R_w)}{\sqrt{\varp(R_w)}}$    & \multicolumn{1}{l|}{0.69}  & 1.87         & \multicolumn{1}{l|}{2.03}          & 2.50         & \multicolumn{1}{l|}{4.96}        & 4.53       \\ \hline
$R_\d$                            & \multicolumn{1}{l|}{-136}          & -105           & \multicolumn{1}{l|}{1260}          & 1222         & \multicolumn{1}{l|}{1351}        & 1287       \\
$\epp(R_\d)$                         & \multicolumn{1}{l|}{30}            & 30           & \multicolumn{1}{l|}{1165}          & 1165         & \multicolumn{1}{l|}{1135}        & 1135       \\
$\varp(R_\d)$                       & \multicolumn{1}{l|}{4062}          & 1210         & \multicolumn{1}{l|}{2781}          & 779         & \multicolumn{1}{l|}{2614}        & 743       \\
$\frac{R_\d-\epp(R_\d)}{\sqrt{\varp(R_\d)}}$ & \multicolumn{1}{l|}{-2.60}          & -3.88         & \multicolumn{1}{l|}{1.80}          & 2.04         & \multicolumn{1}{l|}{4.22}        & 5.57       \\ \hline
\end{tabular}
\label{Rw_Rd_spk}
\end{table}

\subsection{New York City taxi travel pattern}
Finally, we consider daily taxi trips to Central Park from February to April 2014, obtained from the NYC Taxi and Limousine Commission \url{https://www1.
nyc.gov/site/tlc/about/tlc-trip-record-data.page}. The dataset records trip details including pickup and drop-off locations (latitude and longitude) and pickup/drop-off times.

We preprocessed the data similarly to \citep{chu2019asymptotic} by defining the boundaries of New York City as latitudes ranging from 40.577 to 41.5 and longitudes from -74.2 to -73.6. We divided this geographical area into a 30×30 grid with uniform cell sizes. For each cell in the grid, we counted the number of trips originating from that cell and terminating at Central Park each day. The differences between these daily trip count matrices were quantified using the Frobenius norm.
We investigate potential changes in the origins of trips ending in Central Park during the period from February 1st to April 30th.  Given that changes in travel patterns are not uncommon, we kept this sequence short to minimize the impact of the multiple testing problem arising from too many change-points, which is not the primary focus of this paper. 

We applied applied scan statistics with GET and MET on both the robust and standard 5-NNG, as well as e.divisive, to detect change-points. For each method, we identified all change-points in the sequence using binary segmentation with a 0.01 significance level. This process involved continuing the detection on the subsequences $X_1,{\cdots}, X_t$ and $X_{t+1},{\cdots}, X_N$ if $t$ was identified as a significantly change-point.  

\begin{table}[!b]
	\caption{Estimated change-points in New York City taxi data}
\centering	
\begin{tabular}{|r|r|r|r|r|}
		\hline
		  GET on r5-NNG &MET on r5-NNG& GET on 5-MST & MET on 5-NNG & e.divisive\\ 
		\hline
         49, 63&49, 61&NULL&49, 61& NULL \\ \hline
	\end{tabular}
	\label{change-point on New York taxi}
\end{table}

\begin{figure}[!b]
    \centering
    \includegraphics[scale = 0.6]{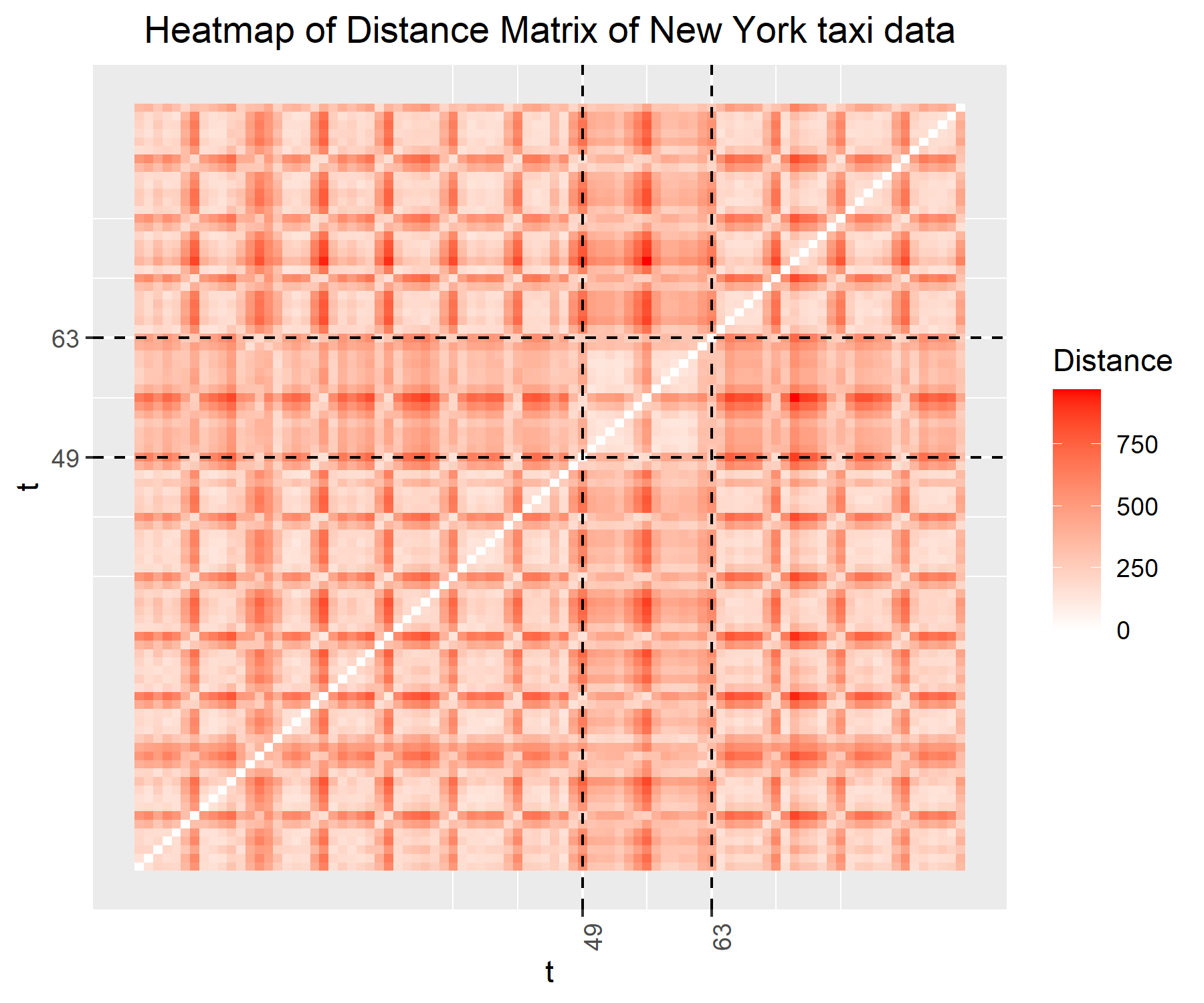}
    \caption{Heatmap of pairwise distances for NYC taxi travel data.}
    \label{heatmap}
\end{figure}

Table \ref{change-point on New York taxi} summarizes the estimated significant change-points for these methods.
Table \ref{change-point on New York taxi} summarizes the estimated change-points from different methods. Both GET and MET with the robust 5-NNG, as well as MET with the standard 5-NNG, identified two change-points: one at 49 (March 21) and another at either 61 (April 2) or 63 (April 4). In contrast, GET on the standard 5-NNG and e.divisive did not detect any change-points. Figure \ref{heatmap} shows a heatmap of pairwise distances, where a clear difference between days 49–63 and the rest of the sequence is evident. These findings suggest that the change-points at 49 and 63 (or 61) are highly plausible. A review of Central Park events in 2014 indicates that this period coincided with the early spring bloom, when visitor activity increases sharply. The robust 5-NNG enabled both GET and MET scan statistics to capture these distributional shifts, whereas GET on the standard 5-NNG failed to do so.

\section{Conclusion and discussion}\label{sec:conclusion}
This paper addresses the challenge of dimensionality effects in graph-based methods, where standard similarity graphs such as the $K$-NNG and $K$-MST are prone to hub formation, leading to loss of power. To mitigate this issue, we proposed a robust graph construction that penalizes the sum of squared node degrees, thereby suppressing hub formation and enhancing the effectiveness of the method.

We also established a theoretical foundation for the proposed approach. Extending the asymptotic framework of \cite{zhu2024limiting} from undirected to directed graphs, we derived sufficient conditions for the validity of the generalized edge-count test on the robust $K$-NNG and proved its consistency under both fixed- and high-dimensional regimes. These results confirm that the proposed construction not only reduces hub effects in practice but also possesses sound asymptotic guarantees.

Extensive numerical studies demonstrated that incorporating the robust graph substantially improves the power of graph-based two-sample tests and offline change-point detection across a wide range of scenarios, including symmetric vs. asymmetric and Gaussian-tailed vs. heavy-tailed distributions. Improvements are especially pronounced when detecting variance or scale differences, where the reduction in variability of the $R_\d$ component enhances sensitivity. Real-data applications to brain imaging, T cell gene expression, and taxi travel patterns further highlight the practical benefits of the robust graph. 

While our focus has been on two-sample testing and offline change-point detection, the robust graph is broadly applicable. In particular, adapting graph-based online change-point detection frameworks remains an important avenue for future research. Current methods are restricted to the standard $K$-NNG \citep{chen2019sequential, chu2022sequential}, and extending them to incorporate robust graphs will require new methodological advances. We leave this as a key direction for future work.

\appendix

\section{Estimated power and detection accuracy in Settings CP2-CP4} \label{plots of change-point study}
\begin{figure}
    \centering
    \includegraphics[width=\textwidth]{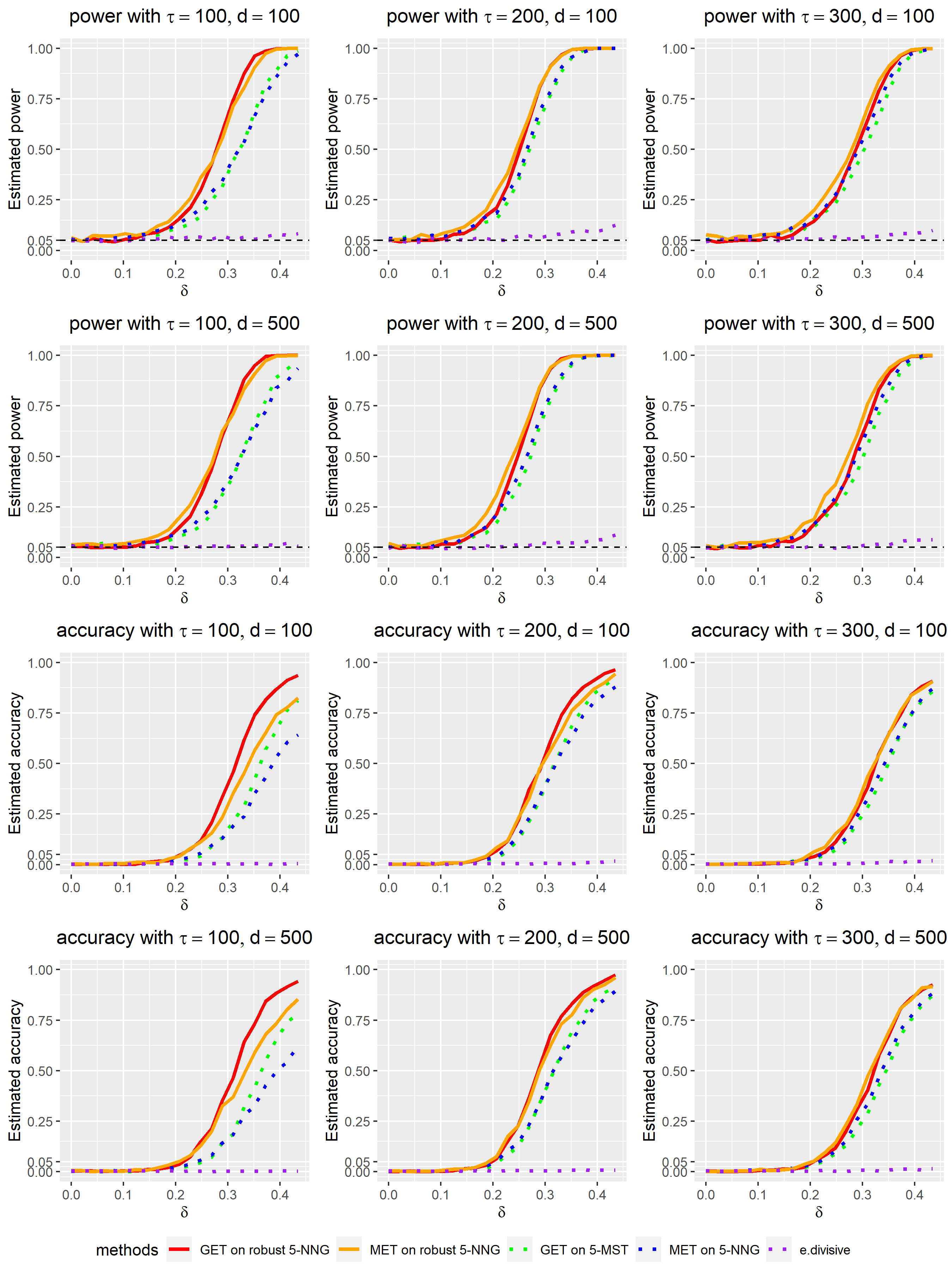}
    \caption{Estimated power and detection accuracy of change-point detection methods under Setting CP2 (covariance change in multivarite Gaussian data).}
    \label{cp_lognormal_diff}
\end{figure}

\begin{figure}
    \centering
    \includegraphics[width=\textwidth]{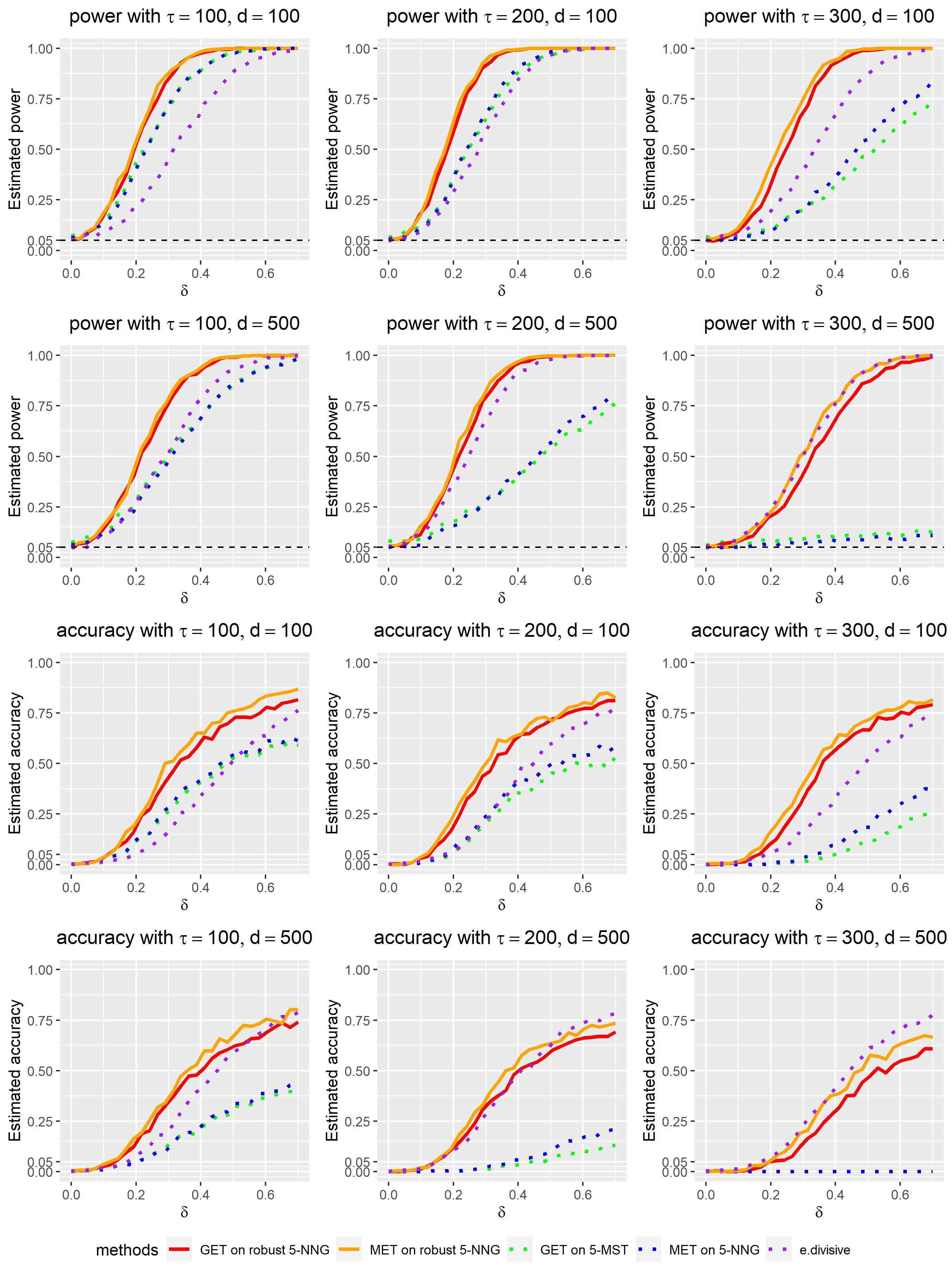}
    \caption{Estimated power and detection accuracy of change-point detection methods under Setting CP3 (location and scale differences in multivariate $t_5$ distribution).}
    \label{cp_t1}
\end{figure}

\begin{figure}
    \centering
    \includegraphics[width=\textwidth]{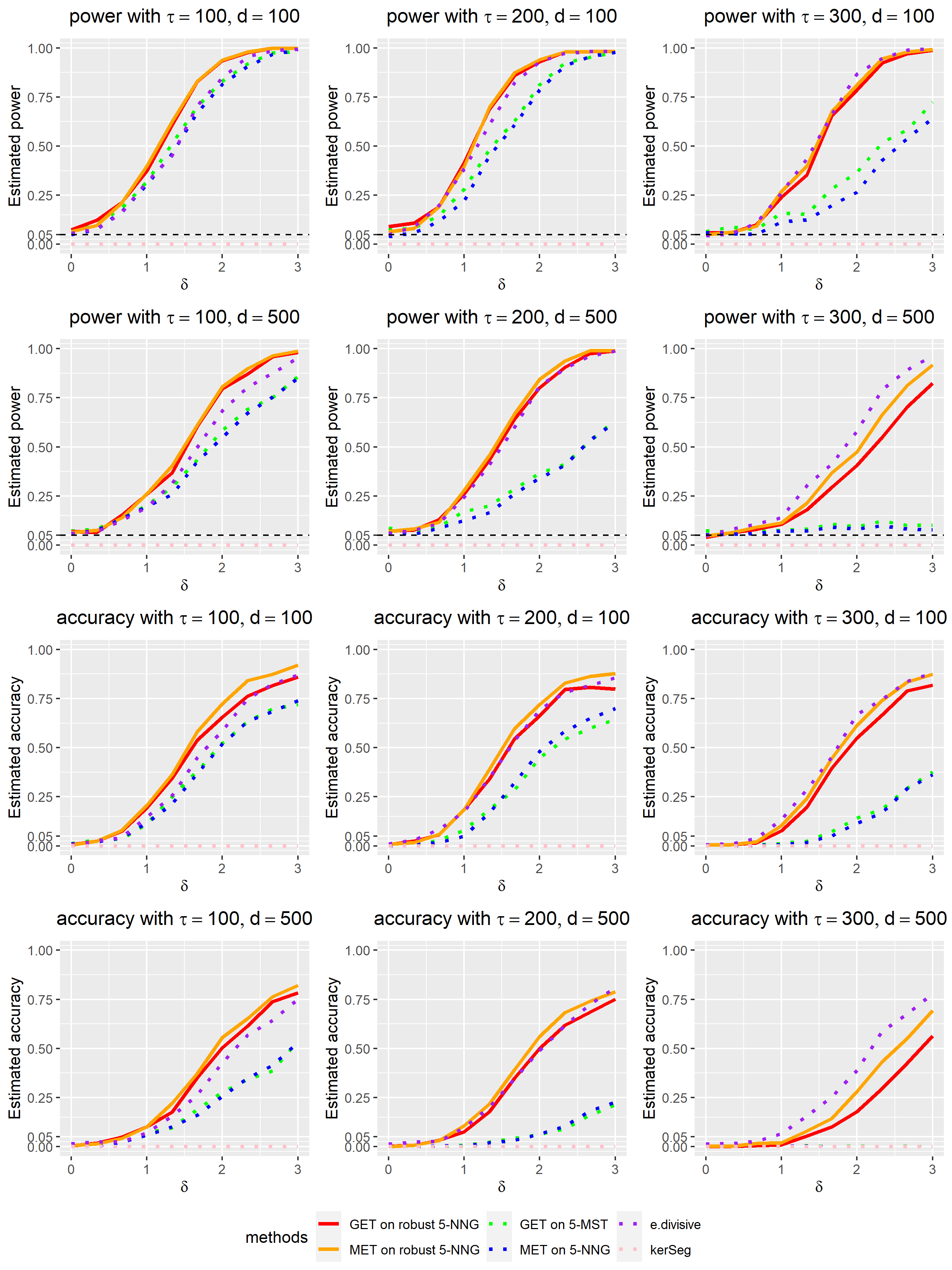}
    \caption{Estimated power and detection accuracy of change-point detection methods under Setting CP4 (location and scale differences in multivariate lognormal distribution).}
    \label{cp_setting4}
\end{figure}

Estimated power and detection accuracy of change-point detection under Setting CP2-CP4 are presented in Fig.~\ref{cp_lognormal_diff}-~\ref{cp_setting4}.

\vskip 0.2in
\bibliography{reference}

\end{document}